\pdfoutput=1

\documentclass[peerreview]{IEEEtran}
\makeatother
\usepackage{graphicx}
\usepackage{amsfonts}
\usepackage{amssymb}
\usepackage{amsmath}
\usepackage{url}
\usepackage{subcaption}
\usepackage{caption}
\expandafter\def\csname ver@subfig.sty\endcsname{}
\usepackage{subfig}
\newcommand{\mtx}[1]{\mathbf{#1}}

\begin{document}
\title{Network Coding as a Service}

\author{
    D\'avid Szab\'o$^{1}$,
    Attila Csoma$^{1}$,
    P\'eter Megyesi$^{1}$,
    Andr\'as Guly\'as$^{1~2}$,
    Frank H.P. Fitzek$^{3~4}$

  \thanks{$^{1}$Budapest University of Technology and Economics, Hungary}
  \thanks{$^{1}$HSNLab, Dept. of Telecommunications and Media Informatics}
  \thanks{$^{2}$MTA-BME Information systems research group}
  \thanks{$^{3}$Technische Universit\"at Dresden, Germany}
  \thanks{$^{4}$5G Lab Germany}
  \thanks{Email: \{szabod,csoma,megyesi,gulyas\}@tmit.bme.hu frank.fitzek@tu-dresden.de
  }

}

\maketitle

\begin{abstract}
Network Coding (NC) shows great potential in various communication scenarios through changing the packet forwarding principles of current networks. It can improve not only throughput, latency, reliability and security but also alleviates the need of coordination in many cases. However, it is still controversial due to widespread misunderstandings on how to exploit the advantages of it. The aim of the paper is to facilitate the usage of NC by $(i)$ explaining how it can improve the performance of the network (regardless the existence of any butterfly in the network), $(ii)$ showing how Software Defined Networking (SDN) can resolve the crucial problems of deployment and orchestration of NC elements, and $(iii)$ providing a prototype architecture with measurement results on the performance of our network coding capable software router implementation compared by fountain codes. 
\end{abstract}


\section{Introduction}
According to the traditional concept of packet switched networks independent data flows may share network devices but the information itself remains separated. NC breaks up this principle as it treats these flows as algebraically combinable information, thereby when two flows $f_1$, $f_2$ enter a node Kirchoff's law doesn't hold any more; the output appears not as $f_1 + f_2$ but $F(f_1,f_2)$.

In the seminal paper \cite{ahlswede2000network}  Ahlswede at al. show a butterfly topology for illustrating that even a simple bitwise \textsc{xor}, i.e. $F(f_1,f_2)=f_1 \textsc{xor} f_2$, used at the proper node of the network can lead to a considerable throughput gain. Unfortunately this nice example also caused confusion as many people mistakenly think that NC can only be used in such a far-fetched, artificial situation. This misunderstanding often appears in literature \cite{medard2012network} and overshadows its intense evolution during the last decade. 

Nowadays NC is not only about butterflies and \textsc{xor} operations but instead creating linear combinations in a distributed way using random number generators, known as ``random linear network code'' (RLNC) \cite{koetter2003algebraic}. RLNC also introduces \emph{recoding}, i.e. to recombine the flows without first decoding them thus fundamentally changes nodes' behaviour since it replaces the \emph{store-and-forward} approach with \emph{compute-and-forward}. This is ground breaking to all other coding strategies that are only end-to-end based (Reed-Solomon, Raptor, etc.) and leads to gain not only for throughput, but for latency, security and complexity as well, furthermore it is feasible for any topology.

However, for the efficient use of RLNC we have to deploy $encoder$, $decoder$ and $recoder$ elements in the network and we have to take care of steering the traffic properly over them. At this point Software Defined Networking with Network Function Virtualization (NFV)  can be the ``door opener'' technologies for RLNC, since they enable to implement RLNC specific features as Virtualized Network Functions (VNFs) \cite{VNFs} that can be connected, or chained, to create communication services. These VNFs then can easily be orchestrated by the control layer of SDN.

The integration of network coding into SDN has alreay been proposed. \cite{nemeth2012towards} and \cite{liu2014ncos} discuss the possibilities of using XOR type network coding through the extension of OpenFlow protocol~\cite{mckeown2008openflow} and functions of switches. In \cite{pavsic2015delay} authors investigate the delay bounds in survivable routing with network coding in SDN. In this paper we extend our prior works \cite{szabotowards,Szabo:2015:TRS:2785956.2790025} that is orthogonal to the aforementioned ones as we investigate the usability of RLNC through NFV in SDNs, that doesn't require to modify existing devices or protocols (e.g. OpenFlow), and also provide a proof of concept.

In the following we give an overview about RLNC and also highlight its features that make it fundamentally different from the generally used block codes (Section \ref{sec:RLNC_description}), then we describe our SDN prototype architecture - with implementation details - that enables the definition, configuration and automated deployment of RLNC specific VNFs (Section \ref{sec:sdn}), finally we provide an extensive comparison about the performance of RLNC and block codes with analytical and measurement results side-by-side (Section \ref{sec:compare}). Section \ref{sec:con} concludes the paper with future work.

\section{Random Linear Network Coding}
\label{sec:RLNC_description}

\subsection{Fundamentals}

RLNC treats data in the form of generations and symbols. A symbol $\vec{s}$ is a vector of Galois Field (GF) elements that represent some data depending on the number of elements $n$ and the size of the GF $f$ according to $|\vec{s}| = n \cdot log_{2}(f) \text{~[byte]}$. For example $8$ elements in $\textsc{GF(2)}$ can represent $1$ byte. A generation $G$ comprises $g$ symbols of size $|\vec{s}|$, so it can represent $g \cdot |\vec{s}|$ bytes of data and it is arranged into a matrix $\mtx{M}=[{s_1}^\intercal,{s_2}^\intercal,...,{s_g}^\intercal]$.

In order to create a coded symbol $\vec{s_c}$ a coding vector $\vec{v}$ is required, that contains a coefficient -- which is an element of GF -- for each symbol in the generation. To encode a new symbol $\vec{s_c}$ from a generation at the source $\mtx{M}$ is multiplied with a randomly generated coding vector $\vec{v}$ of length $g$, $\vec{s_c} = \mtx{M} \cdot {\vec{v}}^\intercal$. In this way we can produce $g+e$ coded symbols, where $e \in \mathbb{Z}$ is the number of extra symbols as RLNC is a rate-less code. 

When a coded symbol is transmitted on the network it is accompanied by its coding vector, and together they form a coded packet $p_c = \{\vec{v},\vec{s_c}\}$. In order to successfully decode a generation the decoder has to receive $g$ linearly independent symbols and coding vectors from that generation. All received symbols are placed in the matrix $\mtx{S_c} = [{\vec{s_c}_1}^\intercal,{\vec{s_c}_2}^\intercal,...,{\vec{s_c}_g}^\intercal]$ and all coding vectors are placed in the matrix $\mtx{V} = [{{\vec{v_1}}^\intercal,{\vec{v_2}}^\intercal,...,{\vec{v_g}}^\intercal}]$. The original data then can be decoded as $\mtx{M} = \mtx{S_c} \cdot \mtx{V}^{-1}$. 

In practice approximately any $g$ coded symbols are enough for a successful decoding of generation $G$. Certainly it is possible to receive a linearly dependent symbol but the chances of this is negligible by using at least GF(8), furthermore, sending another randomly coded symbol is a much looser constraint compared to when no coding is used, where exactly all $g$ unique, original symbols have to be collected.  

However, the most important feature of RLNC is recoding. Any node that received at least $g' \in [2,g]$ linearly independent symbols from a generation and its rank is equal to the rank of $\mtx{V}$, can recode. All received symbols are placed in the matrix $\mtx{S_c} = [{\vec{s_c}_1}^\intercal,{\vec{s_c}_2}^\intercal,...,{\vec{s_c}_{g'}}^\intercal]$  and all coding vectors in the matrix $\mtx{V} = [{{\vec{v_1}}^\intercal,{\vec{v_2}}^\intercal,...,{\vec{v_{g'}}}^\intercal}]$. To recode a symbol these matrices are multiplied with a randomly generated vector $\vec{w}$ of length $g'$, $\vec{v} = \mtx{V} \cdot {\vec{w}}^\intercal$, $\vec{s_c} = \mtx{S_c} \cdot {\vec{w}}^\intercal$. In practice this means that a node that have received more than one symbol can recombine those symbols into recoded symbols, similar to the way coded symbols are constructed at the source, but without having to wait for the whole generation. In other words, a node can change its behaviour from \textit{store-and-forward} to \textit{compute-and-forward}.

\begin{figure*}
\centering
\begin{subfigure}[t]{.32\columnwidth}
\includegraphics[trim=0cm 0cm 0cm 1cm, width=6.5cm]{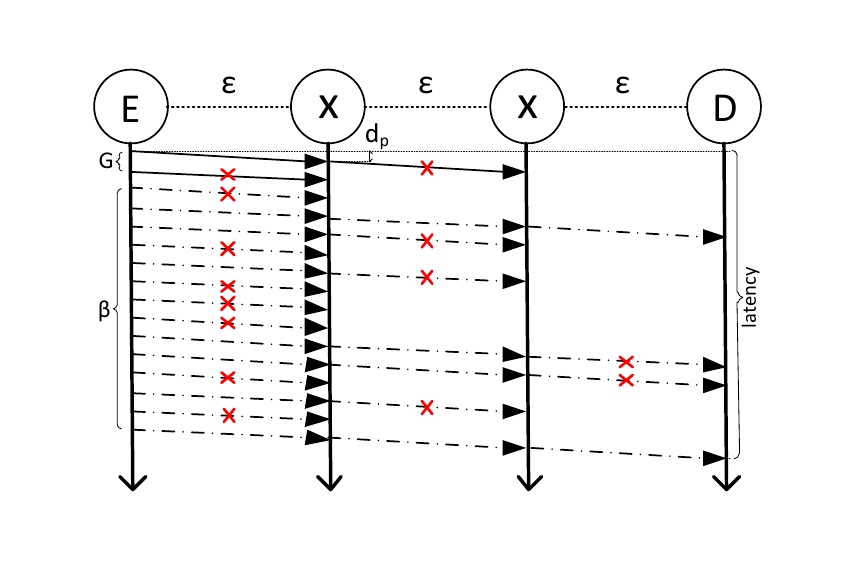}%
\caption{$end-to-end$}
\label{fig:schemes:e2e}
\end{subfigure}
\begin{subfigure}[t]{.32\columnwidth}
\includegraphics[trim=0cm 0cm 0cm 1cm, width=6.5cm]{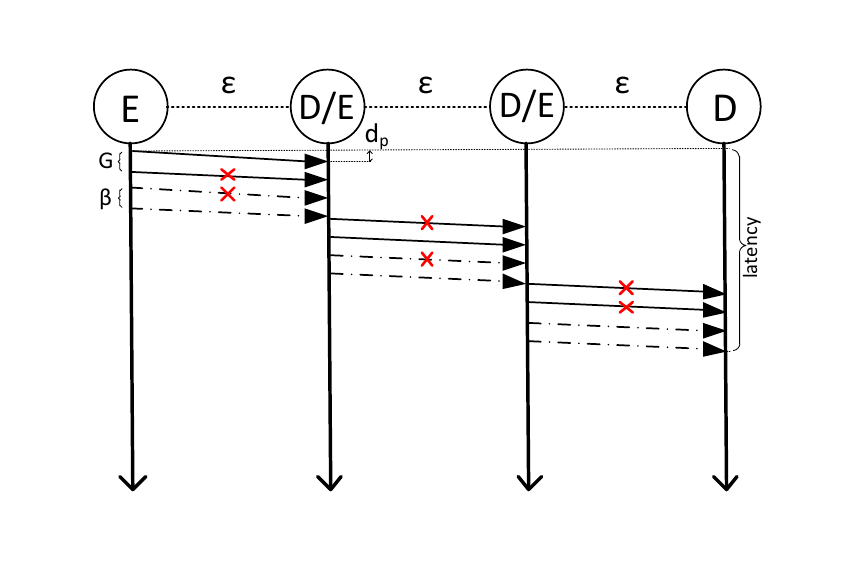}%
\caption{$hop-by-hop$}%
\label{fig:schemes:hbh}
\end{subfigure}
\begin{subfigure}[t]{.32\columnwidth}
\includegraphics[trim=0cm 0cm 0cm 1cm, width=6.5cm]{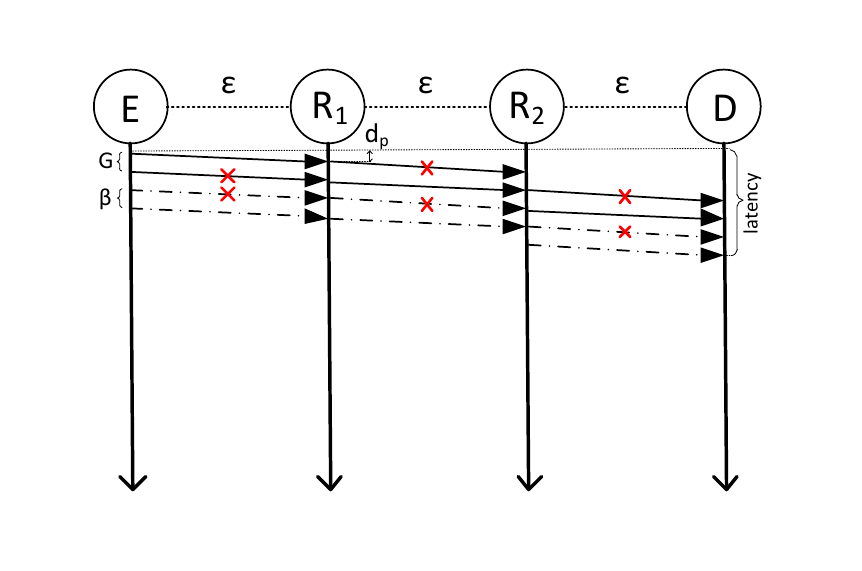}%
\caption{$RLNC$}%
\label{fig:schemes:rlnc}
\end{subfigure}
\caption{Illustration of $end-to-end$, $hop-by-hop$ and $RLNC$ coding schemes for sending a message of 2 packets ($G$) through three hops with 50\% probability loss on each link ($\varepsilon = 0.5$). $\beta$ stands for redundancy (700\%, 50\% and 50\% for E2E, HbH and RLNC, respectively) and $d_p$ for propagation delay.}
\label{fig:schemes}
\end{figure*}

\subsection{Insights of the benefits of RLNC}\label{sec:RLNC_benefits}

In order to shed some more light on this conceptual change we take a closer look on three different coding schemes that are carried out on a \emph{single path} - \emph{multihop} channel. This restriction may seem strange at first glance but we have two good reasons for doing this: $(i)$ packet forwarding on the Internet is mostly carried out in this way, and $(ii)$ we would like to dispel the common misunderstanding that NC cannot be used in any other case than multicast.

Accordingly we assume an encoder $E$ that delivers a message comprises $G$ packets to a decoder $D$. Along the path packets forwarded by multiple relay nodes ($X$s) and we also assume error prone links with loss probability $0 \leq \varepsilon \leq 1$ (Fig.~\ref{fig:schemes}). We consider the following three cases:

\noindent \emph{Block codes in end-to-end manner (E2E):} In this scheme encoding and decoding are performed only once by $E$ and $D$, respectively. The relay nodes only \emph{store-and-forward} each successfully received packet, implying that $E$ should emit enough amount of extra packets for the whole channel to compensate the loss and to make sure that $D$ can reconstruct the message (Fig.~\ref{fig:schemes:e2e}). This eventuates unnecessary traffic loads on the links closer to $E$, which is particularly painful when losses occur only on links far from $E$, and also increase latency. 

\noindent \emph{Block codes in hop-by-hop manner (HbH):} In this scheme we assume that relay nodes can also perform encoding and decoding that enables to generate extra packets per hop in a distributed way (Fig.~\ref{fig:schemes:hbh}). This unburdens the network from the unnecessary packets but also infuse extra latency as every relay has to wait to start encoding until the full message is decoded.

\noindent \emph{Random Linear Network Coding:} RLNC enables for relays to perform recoding, that is to create a coded packet from the received ones without decoding them first (Fig.~\ref{fig:schemes:rlnc}). This is far less complex than decode/encode procedure, hence eventuates immediate forwarding, furthermore, it is completely transparent, so decoding doesn't require extra effort. This greatly reduces latency and as it is carried out per hop requires the same number of packets as the HbH scheme. However, in order to use RLNC efficiently, first we have to deploy elements with $encoding$, $decoding$ and $recoding$ capabilities and we have to take care of steering the traffic over them. 

In Section \ref{sec:compare} we provide an extensive comparison of RLNC and block codes supported both analytical and measurement results, but as the measurements were carried out on our SDN architecture first we show how SDN and virtualization can facilitate the integration process of RLNC in a seamless way. 

\section{SDN prototype architecture}
\label{sec:sdn}

Here we give a brief introduction about SDN and Network Function Virtualization (NFV), followed by the implementation details of RLNC capable elements, finally we introduce the architecture which enables the definition, configuration and automated deployment of these novel elements implementing code centric operation in the network.

\subsection{SDN and virtualization}
The basic concept of SDN is to enable network innovation - realizing new capabilities and addressing persistent problems with networking -, which is almost impossible nowadays. The problem lies in the distributed and heterogen nature of current networks.  There are closed, vendor specific hardwares, softwares and firmwares across the network managed by distributed control functions through different interfaces. This leads to difficult and extensive design and operation. 

SDN aims to change this by creating well defined abstractions of different network layers, that each has its proper functionality and interfaces. In this sense the idea is somewhat similar to the ISO/OSI conception, but instead of individual devices it concerns the network as a whole with the following features: 
$(i)$ maintains the separation of the data and control planes, 
$(ii)$ uses uniform vendor-agnostic interface -- one of the most commonly used is OpenFlow~\cite{mckeown2008openflow} -- between control and data planes, 
$(iii)$ treats the control plane in a logically, centralized way that is realized using a network operating system that constructs and presents a logical map of the entire network to services or control applications implemented on top of it, and 
$(iv)$ slices and virtualises the underlying physical network. 

During its two decades evolution \cite{feamster2013road} the concept of SDN inspired several novel technologies and turned out that it synergizes well with many existing ones. One that really shines out amongst them is Network Function Virtualization (NFV). NFV enables to implement any hardware middleboxes (a network device that manipulates traffic, e.g. firewall, network address translator, etc.) into software by creating Virtualized Network Functions (VNFs). These VNFs has exactly the same functionality but they don't require specific hardware and hundreds of them can be installed - even remotely - on a single device. This offers nice synergy with SDN as the control layer is capable to orchestrate the installation, configuration and traffic steering over VNFs in an automated way.

Such orchestration and steering is in perfect agreement with the current practice of Internet Service Providers (ISPs) that services are implemented in the form of appropriately concatenated middleboxes, also known as \emph{service chains} \cite{john2013research}. However, today’s service chains are usually built around special purpose networking hardware elements, configuring and operating these chains is a highly non-trivial task which still requires human interaction. SDN and virtualization can be a promising way out of this managerial trap as they enable flexible and automated deployment of service chains comprising our RLNC capable VNFs.

\subsection{Implementation of RLNC Software Router Prototype}
\label{sec:click}

Since VNFs are run on NFV platforms the first design step is to choose one from the many existing ones (\cite{180672},\cite{Sherry2342359},\cite{CiscoImp},\cite{Vyatta}). Our design choice was ClickOS~\cite{martins2014clickos}, as ClickOS virtual machines are extremely small (5 Mb), can boot very quickly (about 30 ms), add small delay (around 45 $\mu$s) and hundreds of them can run concurrently with a throughput around 10 Gb/s. ClickOS requires VNFs created by the Click modular router platform~\cite{click}, which enables to built custom routers by creating configurations pieced together from built-in or own-developed elements that implement atomic functionality like packet classifying, scheduling, queuing etc. Using the Kodo~\cite{pedersen2011kodo} network coding software library we have developed the RLNC encoder, recoder and decoder Click elements and using built-in modules we have created fully-fledged compute and forward software routers.

\begin{figure*}[t!]
 \centering
 \captionsetup{justification=centering}
 \includegraphics[trim=0cm 0cm 0cm 1cm, width=0.59\textwidth]{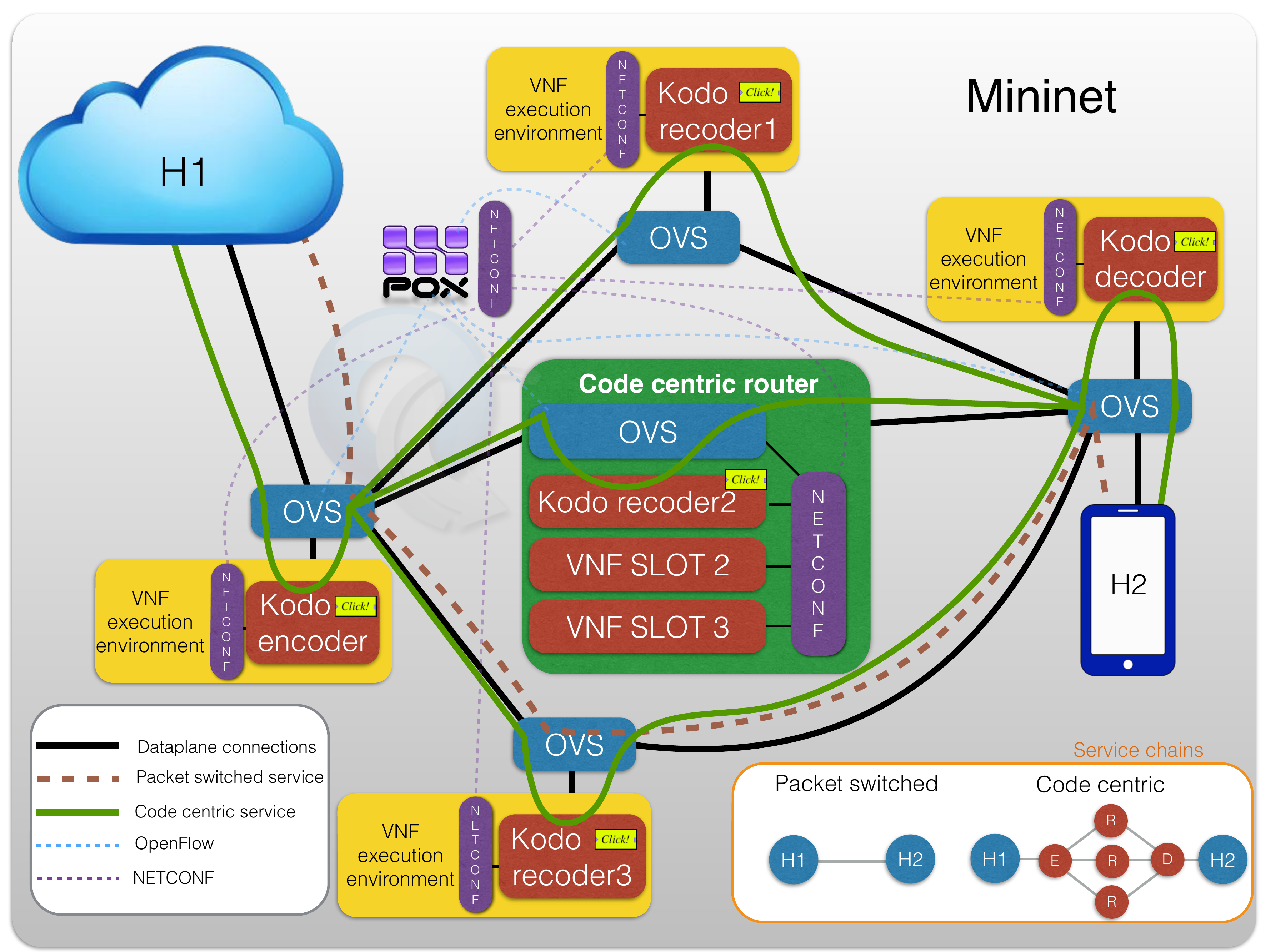}
 \caption{The prototype architecture.}
 \label{fig:escape}
\end{figure*}

The Click interpreter reads configurations written in a Click-specific language. These config files describe a directed graph with \emph{elements} at vertices and \emph{edges} specifying possible paths for the packets within the router. The behaviour of the elements are given by C++ code.  At the code level each element is a subclass of the $Element$ class, which has around 20 virtual functions and most subclasses have to override only up to six of them. The most simple subclass implementation, the $NullElement$, contains only 8 lines of code and overrides five $Element$ class functions. Similarly, we have implemented our code-centric elements which encode, recode and decode the incoming packets using Kodo.

Kodo offers a number of different erasure correcting codes of which we chose Full Random Linear Network Code (Full RLNC), as it is one of the most common RLNC variants and provides several of the advantages that RLNCs have over traditional erasure correcting codes.  Accordingly our three custom Click elements are $FullRLNCEncoder$, $FullRLNCDecoder$ and $FullRLNCRecoder$. All of them take around 120-150 lines of C++ code (with no particular optimization) and override seven virtual functions of the $Element$ class. They have one input and one output port and the coding parameters can be tuned through input arguments: \textsc{symbols, symbol\_size, gf\_size, extra}. The \textsc{symbols} argument stands for the generation size and tells the maximal number of symbols that can be combined into a coded symbol by the encoder. Increasing the generation size also increases the decoding delay, since the decoder has to receive at least \textsc{symbols} number of packets to be able to decode the whole generation. \textsc{symbol\_size} represents the size of each symbol in bytes. Increasing this eventuates increased coding complexity. So the \textsc{symbols}$*$\textsc{symbol\_size} product should be considered carefully and large data typically sent through multiple generations. The \textsc{gf\_size} argument stands for the size of the Galois Field, which has influence on the probability that an encoded packet doesn't carry any useful information. Finally the \textsc{extra} parameter represents the ratio of redundancy, in other words tells how many extra packet should be generated. This parameter is required only for the encoder and the recoder.

Since other built-in Click elements can preprocess UDP\footnote{Our current implementation can process UDP packets only. The handling of TCP flows is in our future work list. } packets (i.e. strip IP and UDP headers) the general behaviour of our code-centric elements is quite simple: After a packet arrives extract the payload, encode/recode/decode it by calling the proper functions of Kodo, update IP and UDP headers (size and checksum fields as we slightly increase the packet size by adding the coding coefficients) and forward the packet. In this way our router configurations implement the compute and forward router as a VNF that performs code-centric operations on the packet going through and can easily be deployed into SDN environment.

\subsection{Architecture}

For showing the seamless integration of network coding and SDN we have built a prototype of the code centric network architecture. In a real network it would be a practical choice for operators to place NC encoder and decoder elements as close to the edge of the network as possible, while recoders operate in the most efficient way at intermediate nodes where they can aggreagate traffic flows. To test our proof-of-concept we created a smaller network to model a similar topology where every packet between two users traverses an encoder a recoder and a decoder (in this order). To simulate such network we implemented our architecture in the ESCAPE prototyping environment. Well detailed description about it can be found in~\cite{Csoma:2014:EES:2619239.2631448}, however, here we recollect the most important information  of the components.

ESCAPE is capable to simulate OpenFlow networks combining Mininet network virtualizer~\cite{mininet-short} using Open vSwitch (OVS)~\cite{pfaff2009extending} instances and a POX netowork control prototyping software~\cite{POX}, which contains a steering module handling the flow tables according to the configuration of the running service chains. ESCAPE is designed in such way that it can initiate Mininet virtual containers which capable to run binaries or source codes written in Click language. To use these containers as VNFs ESCAPE configure OVS elements with POX in such way that all traffic between end users traverse them. Typically after an initial deployment process OVS configuration remains the same during the simulation. Since the controller can automatically deploy VNFs implemented in Click and our software router in Sec.~\ref{sec:click} is implemented in the Click platform too, our work here was to translate our Click configurations according to the templates used by ESCAPE and install all software required to run ESCAPE and Kodo. Note that the traffic steering in a real network with multiple users, diverse requirements and huge amount of independent flows raises several open questions. However, in this paper we restricted ourselves only to provide a proof of concept implementation which requires a minimalistic network with a few traffic flows.

Fig.~\ref{fig:escape} presents a scenario of our code-centric prototype. We build an OpenFlow 1.0 network of three Open vSwitch instances (OVS), two hosts (H1/H2) and three VNF containers. These containers (yellow boxes) are advanced Mininet hosts which can start a given VNF process. This solution stands for the case when we run VNFs outside the routers e.g. in a near OpenStack \cite{sefraoui2012openstack} data center. In these containers we deployed our compute and forward software router configured to encode/recode/decode packets similarly to our scenarios in Fig.~\ref{fig:schemes} for implementing our coding schemes (E2E, HbH and RLNC). This solution may slightly increase latency but provides the possibility of scaling out in the presence of massive network loads combined with more complex coding schemes (e.g. more efficient random linear codes using larger GF field size). 

Alternatively, we have also built a visionary prototype of the code centric router (green box). This router consists of a standard OVS (no modifications in OpenFlow) instance but has the capability to execute VNFs. Using code centric routers adds less delay but there is no way to scale out beyond the router's hardware resources. The POX orchestrator module receives service chains as inputs (can be given by GUI), configure the switches and start the appropriate VNFs accordingly.

\section{Comparing RLNC with block codes}
\label{sec:compare}

A comprehensive analysis can be found in \cite{dikaliotis2009delay} and \cite{pakzad2005coding} about the performance of difference coding schemes including complexity, delay, memory requirement, achievable rate, and adaptability. However, in order to facilitate understanding we recall the most important claims on the number of sent packets and latency adjusted to the communication scenarios described in Section~\ref{sec:RLNC_benefits} and provide  measurements results for a wide range of parameter settings side-by-side.

For the measurements we realized all the three scenarios as service chains in our prototype architecture and besides the properties of the links (erasure probability and bitrate) we varied the number of hops, packet size and coding generation as well (Table~\ref{tbl:params}).

\begin{table}[h]
\centering
\begin{tabular}{|l|c|}
\hline
\textbf{Parameter}             & \textbf{Values}      			\\ 
\hline
erasure probability $\epsilon$ & 10\%, 20\%, 30\%, 40\%, 50\%		\\
packet size $L$                & 250 B, 500 B, 750 B, 1000 B, 1450 B 	\\
generation size $G$            & 16, 32, 64, 128       			\\
number of hops $H$             & 2, 3, 4, 5, 6, 7      			\\
channel rate                   & 0.25, 0.5, 1, 2, 4, 8 Mbps \\ \hline
\end{tabular}
\caption{The parameter set for the measurements. }
\label{tbl:params}
\end{table}

During the analysis we assume a \emph{single path} - \emph{multihop} channel (described in Section~\ref{sec:RLNC_description}), where the encoder $E$ delivers a message of $G$ packets through $H$ number of links to a decoder $D$, we also assume error prone links with loss probability $0 \leq \varepsilon \leq 1$.

\subsection{Number of Sent Packets}

Now we calculate the overall number of sent packets required $D$ to successfully decode the message. In the case of E2E this is the sum of packets sent per hop - comprises extra packets for compensating losses - on the rest of the channel, which depends on generation size $G$, number of hops $H$ and loss probability $\varepsilon$: 
\begin{equation}
  P_{E2E}=\sum_{h=1}^H G \prod_{i=h}^H \frac{1}{1-\epsilon_i}
\end{equation}

For HbH and RLNC it is again the sum of the packets sent individually but here losses have only local impact due to the decode/encode procedure carried out at each hop: 

\begin{figure}[!t]
  \begin{center}
    \includegraphics[trim=1cm 1cm 1cm 2cm, width=0.4\textwidth]{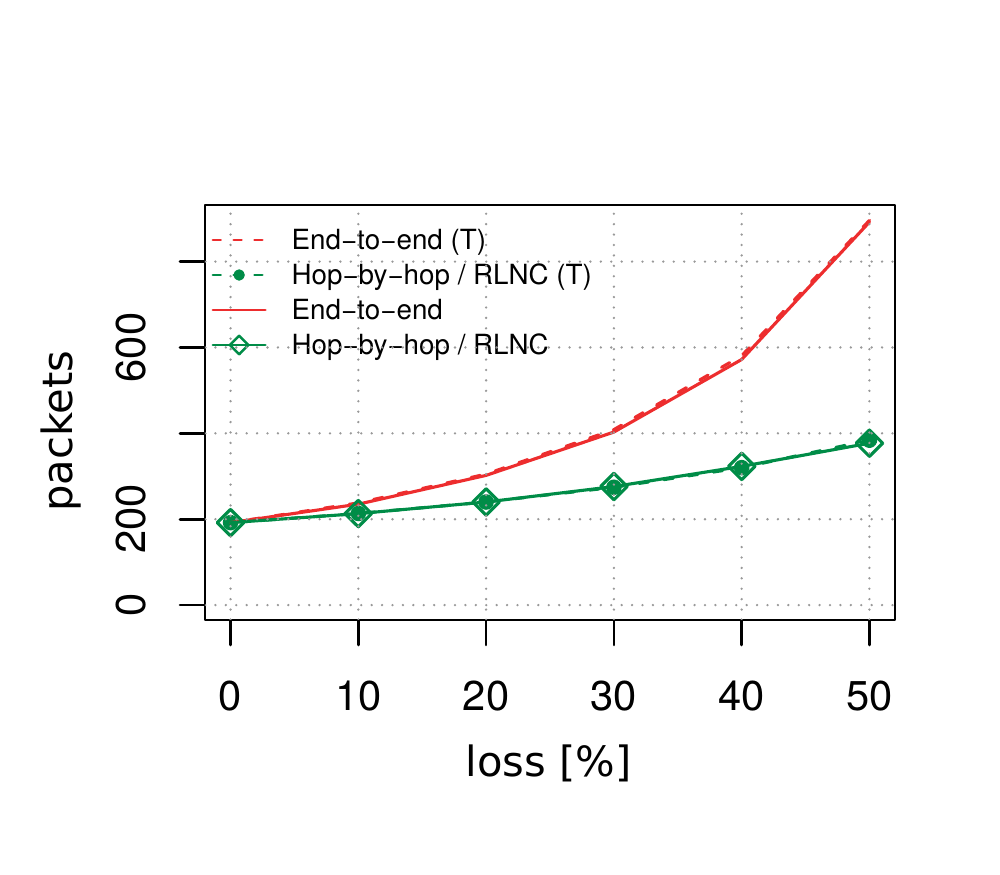}
  \end{center}
  \caption{Number of overall packets conveyed in the network versus
    channel loss probability (Packets 64 - Size 250 B - Hops 3)}
  \label{fig:theory_vs_sim:packets}
\end{figure}

\begin{equation}
P_{HbH}=P_{RLNC}=G \cdot \sum_{i=h}^H \frac{1}{1-\epsilon_i}
\end{equation}

In Fig.~\ref{fig:theory_vs_sim:packets} the number of overall packets conveyed in a three hop communication network versus the erasure probability per link for the three forwarding schemes is given for the theoretical (indicated by (T) in the figure) and for the measurement results. The figure show that the results of the theory and measurements are consistent with each other.  HbH and RLNC use the same amount of packets, and while they increase the number of packets linearly with the loss probability, the E2E approach increases exponentially.

\subsection{Latency}

Based on packet numbers we can calculate the time required for decoding the message successfully. In other words, we are interested in the time that takes to deliver all packets of $G$ from $E$ to $D$ and we also calculate with an inter packet time $\tau_P$ which is the multiplicative inverse of the packet sanding rate and link delay $\tau_L$ that each packet suffers during forwarding (we assume the same dealy for every $H$ links).

In the case of E2E this is the sum of packets emitted at first hop - comprises extra packets as well - and an extra delay per hop (since packets are forwarded immediately and in parallel at each hop):

\begin{figure}[!t]
  \begin{center}
    \includegraphics[trim=1cm 1cm 1cm 2cm, width=0.4\textwidth]{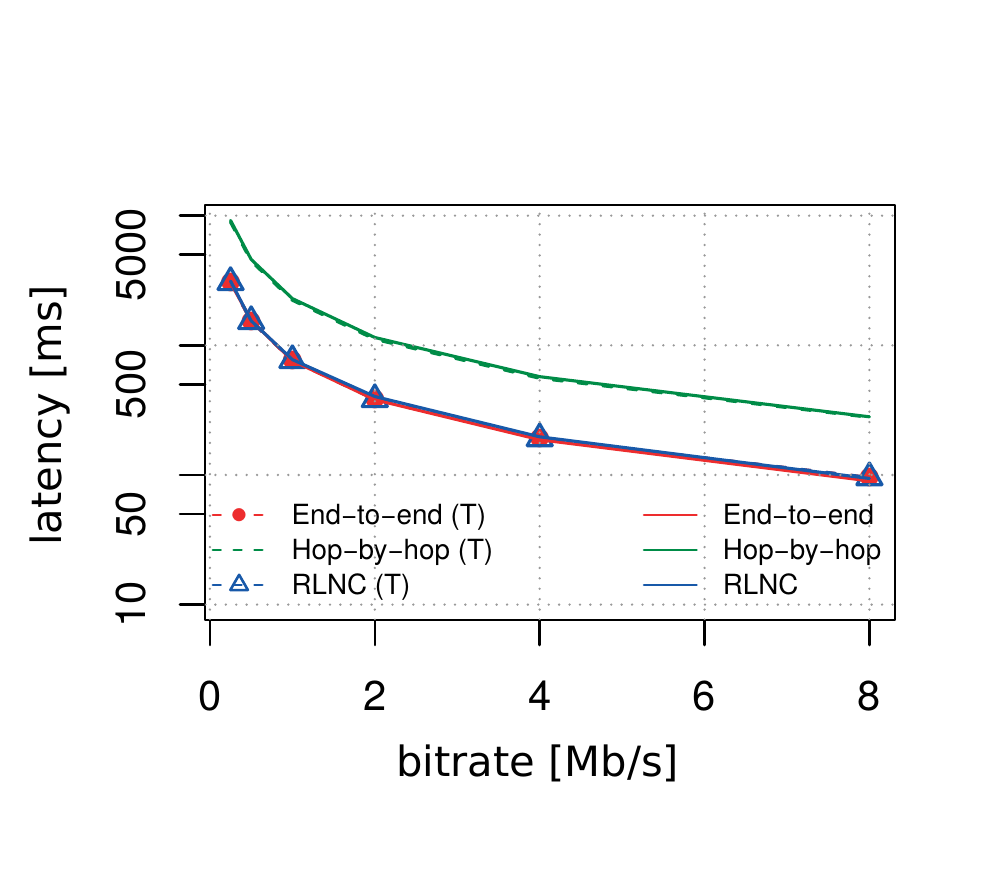}
  \end{center}
  \caption{Latency in the network versus channel rate for three coding 
    approaches and no losses (Packets 64 - Size 1450 B - \textbf{Loss 
    0\%} - Hops 3)}
  \label{fig:theory_vs_sim:latency:zero}
\end{figure}

\setcounter{figure}{5}
\begin{figure*}[!b]
\centering
\begin{subfigure}{.32\columnwidth}
\includegraphics[trim=0cm 1.5cm 0cm 2cm, width=5.5cm]{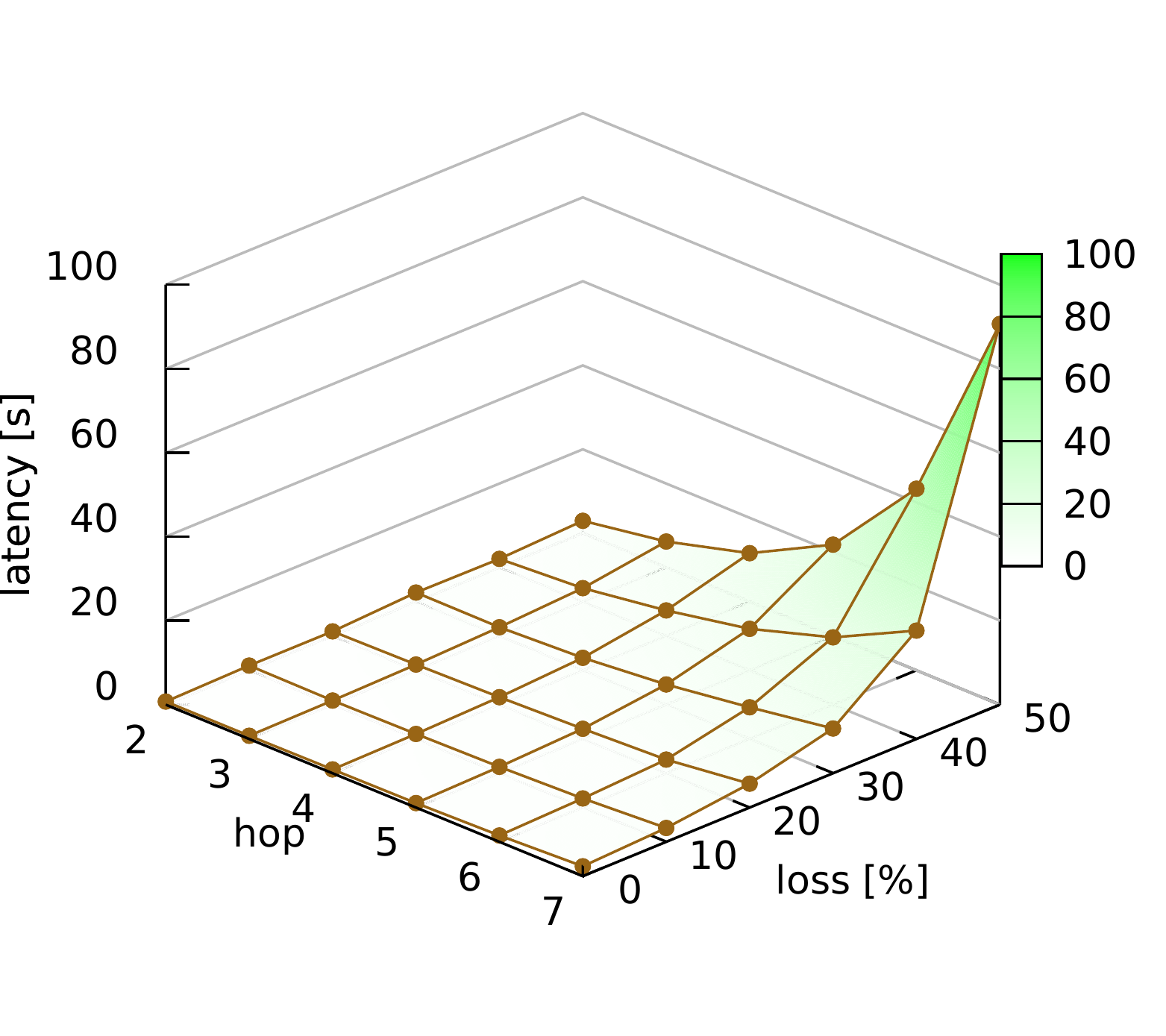}%
\caption{$end-to-end$}
\label{fig:3D_losses:e2e}
\end{subfigure}
\begin{subfigure}{.32\columnwidth}
\includegraphics[trim=0cm 1.5cm 0cm 2cm, width=5.5cm]{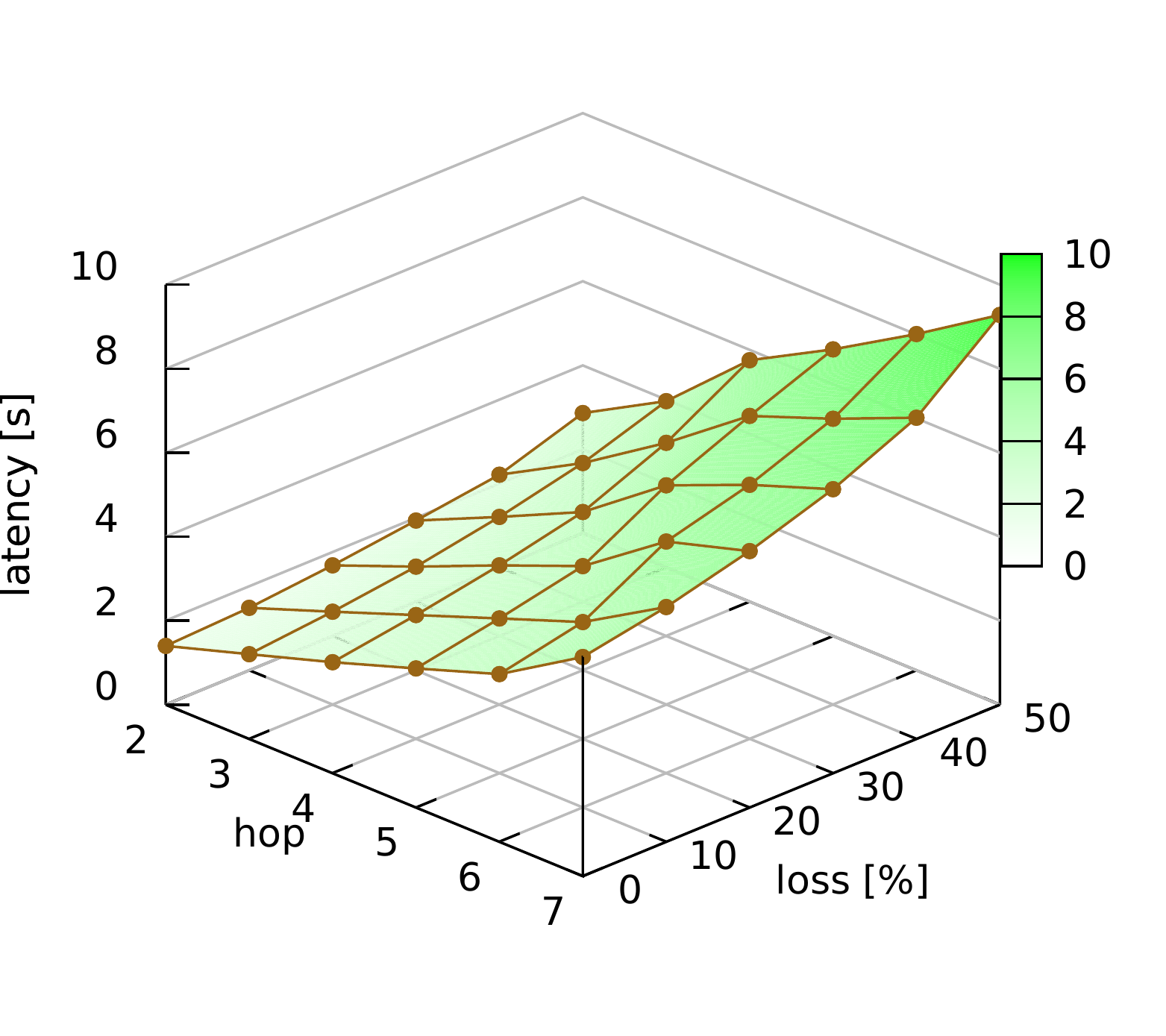}%
\caption{$hop-by-hop$}%
\label{fig:3D_losses:hbh}
\end{subfigure}
\begin{subfigure}{.32\columnwidth}
\includegraphics[trim=0cm 1.5cm 0cm 2cm, 
width=5.5cm]{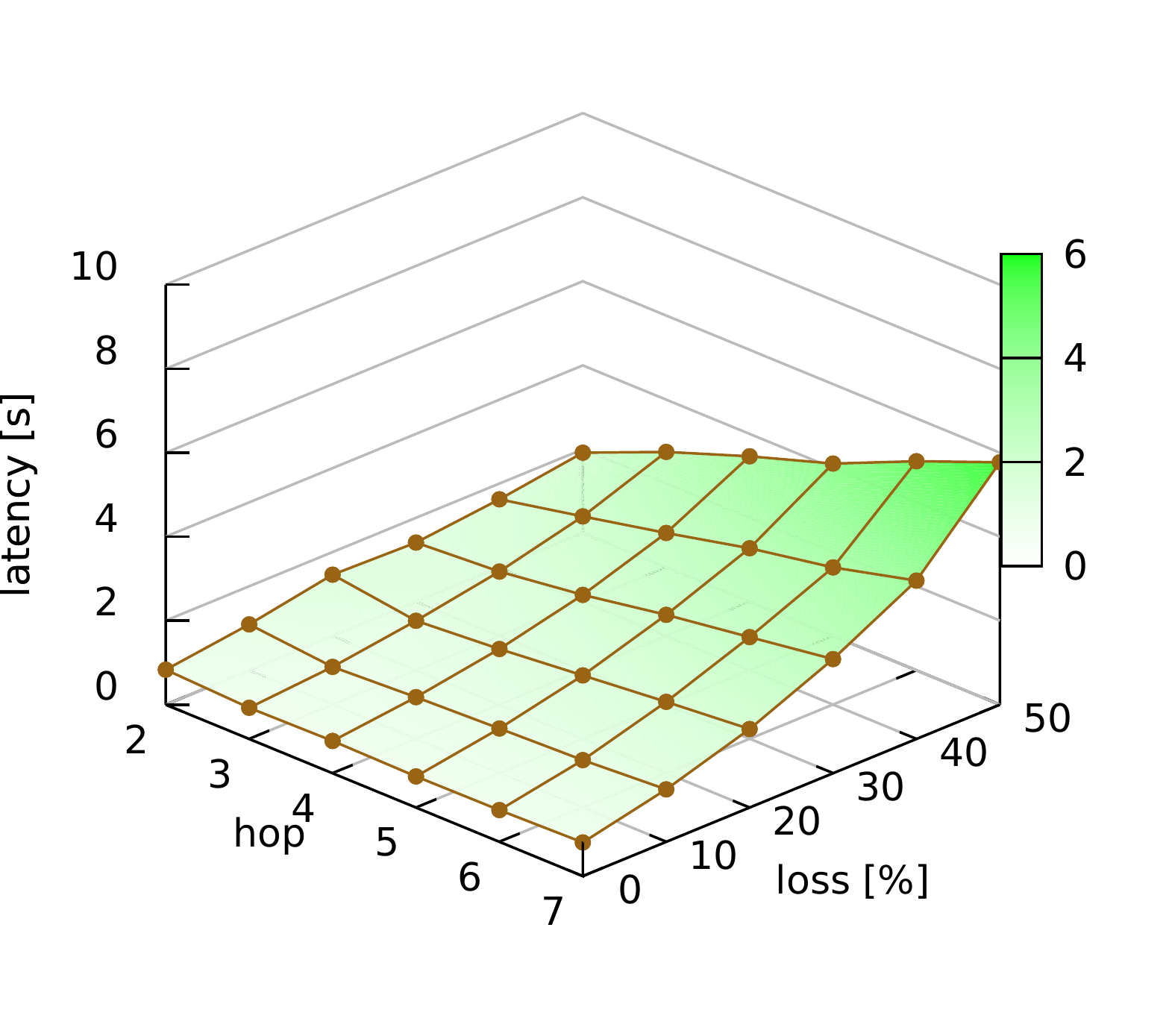}%
\caption{$RLNC$}%
\label{fig:3D_losses:rlnc}
\end{subfigure}
\caption{Latency for the three transmission schemes depending on number of hops 
and loss (Packets 64 - Size 250 B - Bitrate 0.25 Mb/s).}
\label{fig:3D_losses}
\end{figure*}

\begin{equation}
D_{E2E}=G \cdot \prod_{i=1}^H \frac{1}{1-\epsilon_i} \cdot \tau_P + H \cdot \tau_L
\end{equation}

For HbH this is the time for the first hop multiplied by the number of hops as every node has to perform decoding/encoding before forwarding even the first packet:

\begin{equation}
D_{HbH}= G \cdot \tau_P \cdot \sum_{i=1}^H \frac{1}{1-\epsilon_i} + H \cdot \tau_L
\label{delay:hbh}
\end{equation}

The case of RLNC scheme comprises the best part from both E2E and HbH, since packets are forwarded immediately and in parallel with the same number of packets per hop as in HbH:

\begin{equation}
D_{RLNC}=G \cdot \tau_P \cdot \frac{1}{1-\displaystyle\max_{1 \leq i \leq H} \epsilon_i} + H \cdot \tau_L
\label{delay:rlnc}
\end{equation}

\setcounter{figure}{4}
\begin{figure}[!t]
  \begin{center}
    \includegraphics[trim=1cm 1cm 1cm 2cm, width=0.4\textwidth]{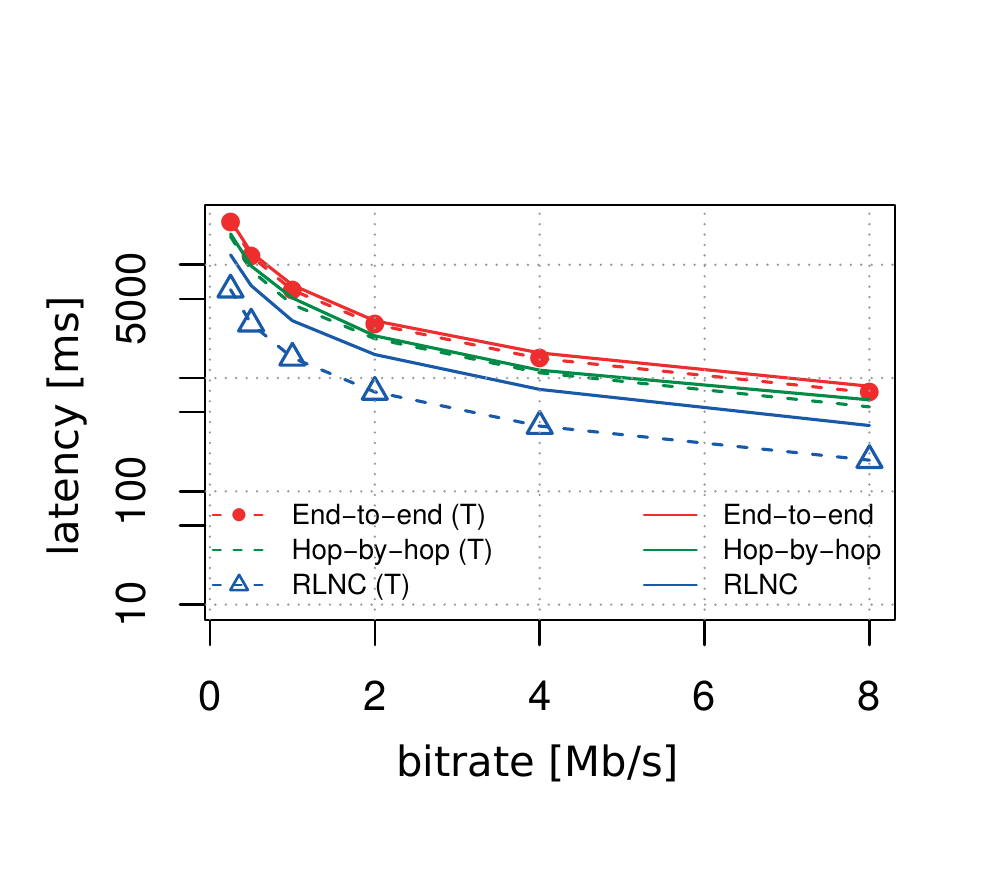}
  \end{center}
  \caption{Latency in the network versus channel rate for three coding
    approaches and high losses (Packets 64 - Size 1450 B -
    \textbf{Loss 50\%} - Hops 3)}
  \label{fig:theory_vs_sim:latency:fifty}
\end{figure}
\setcounter{figure}{6}

Fig.~\ref{fig:theory_vs_sim:latency:zero} presents the latency in a three hop communication network versus channel rate without any losses for the three coding approaches. If there are no losses E2E and RLNC have the same latency values (since no extra packets are required to send, which would slow E2E), while the HbH ends up in higher latency values because each intermediate node needs to wait for all packet of $G$. The gain of RLNC over HbH remains constant for the higher values which means that the ratio of latency is independent from the bandwidth. Roughly, the ratio of the gain in latency equals to $H$, when $G$ is significantly higher than $H$. 

The latency results change a lot if the channel is error prone as given in Fig.~\ref{fig:theory_vs_sim:latency:fifty} with an error probability of $50\%$. Now the advantage of RLNC over the other two schemes becomes evident and E2E is now even worse than HbH. After having a look again at Fig.~\ref{fig:schemes} this is not so surprising, since E2E have to send through all redundancy on the whole channel. While HbH can unburden the network, as redundancy have to be sent per hop, there the \textit{store and forward} behaviour increases latency.

However, in Fig.~\ref{fig:theory_vs_sim:latency:fifty} it can be observed that while measurement follows theory well for E2E and HbH we got much higher values for RLNC. In~\cite{dikaliotis2009delay} authors already proved that in the case of a two-hop network with identical links the delay function grows as $\sqrt{n}$ thus it does not follows the original formula in Eq.~\ref{delay:rlnc}. We investigated this phenomenon further and discuss the reasons in Appendix \ref{appendix}. Suffice it to say for now that RLNC performs a bit worse as theory would suggest when the error probabilities of the links are similar.

In the followings we extend the measurements for a wide variation of parameters and based on the fact that RLNC performs worse than expected when error probabilities are close to each other we set the values accordingly in order to show that even in this case RLNC outperforms the other coding schemes.

\begin{figure}[!b]
  \centering
  \begin{subfigure}{.4\columnwidth}
    \includegraphics[height=.65\linewidth]{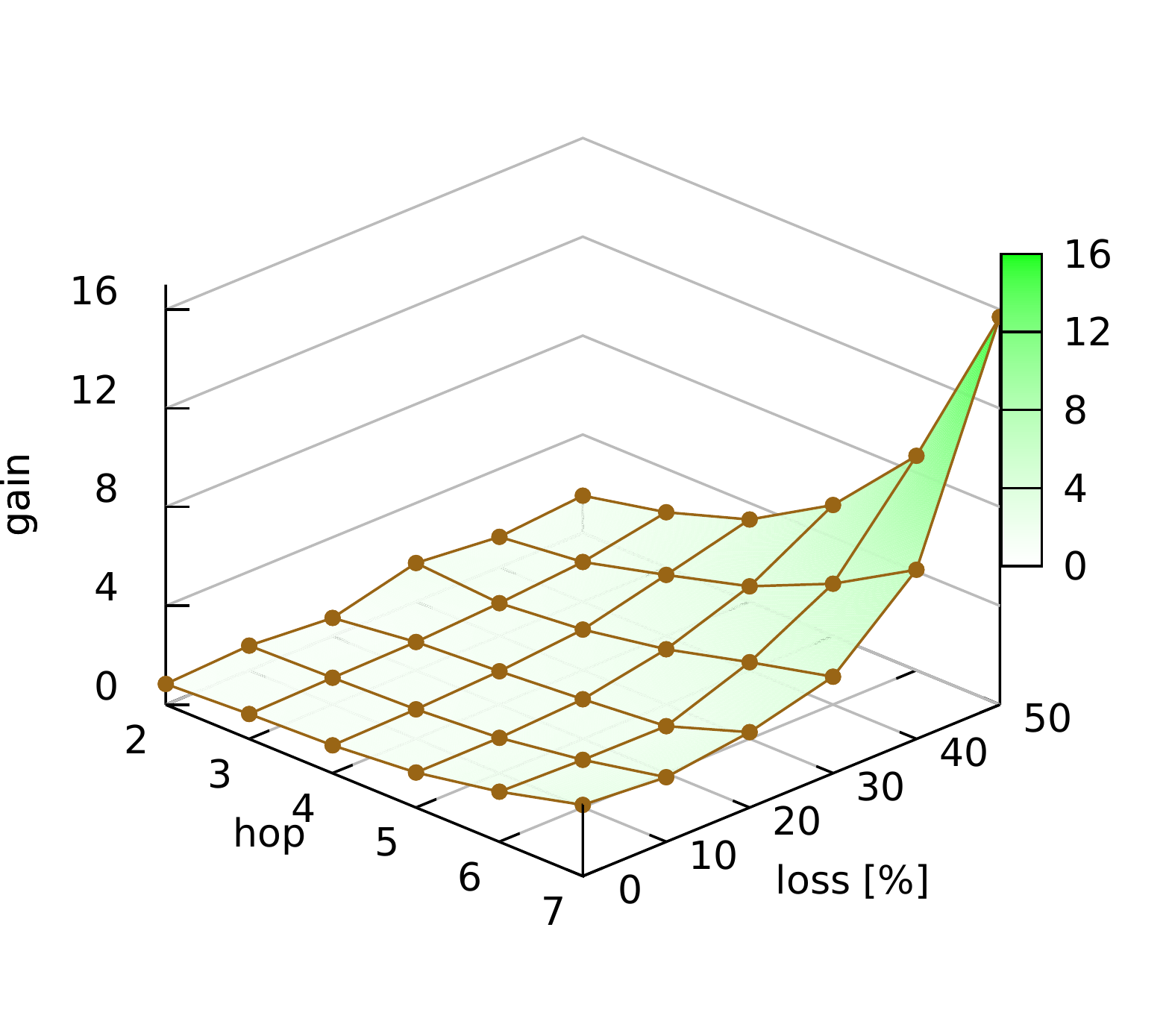}%
    \caption{Gain of RLNC over $E2E$}
    \label{fig:3D_gain:e2e}
  \end{subfigure} 
  \begin{subfigure}{.4\columnwidth}
    \includegraphics[height=.65\linewidth]{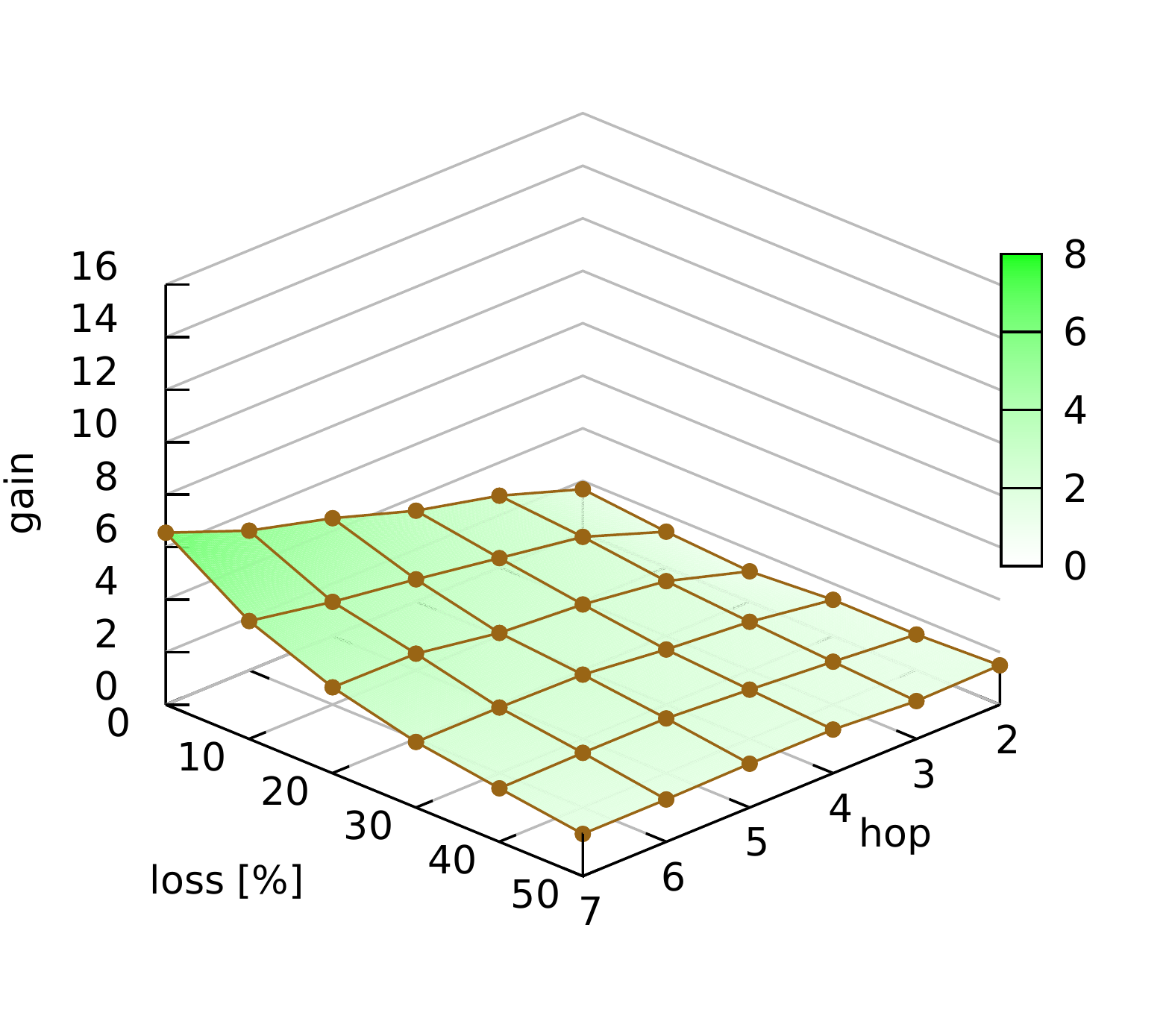}%
    \caption{Gain of RLNC over $HbH$}%
    \label{fig:3D_gain:hbh}
  \end{subfigure}
  \caption{Gains for the three transmission schemes (Packets 64 - Size
    250 B - Bitrate 0.25 Mb/s).}
  \label{fig:3D_gain}
\end{figure}

\begin{figure*}[!t]
  \centering
  \begin{subfigure}{.32\columnwidth}
    \includegraphics[trim=0cm 1.5cm 0cm 2cm,
    width=5.5cm]{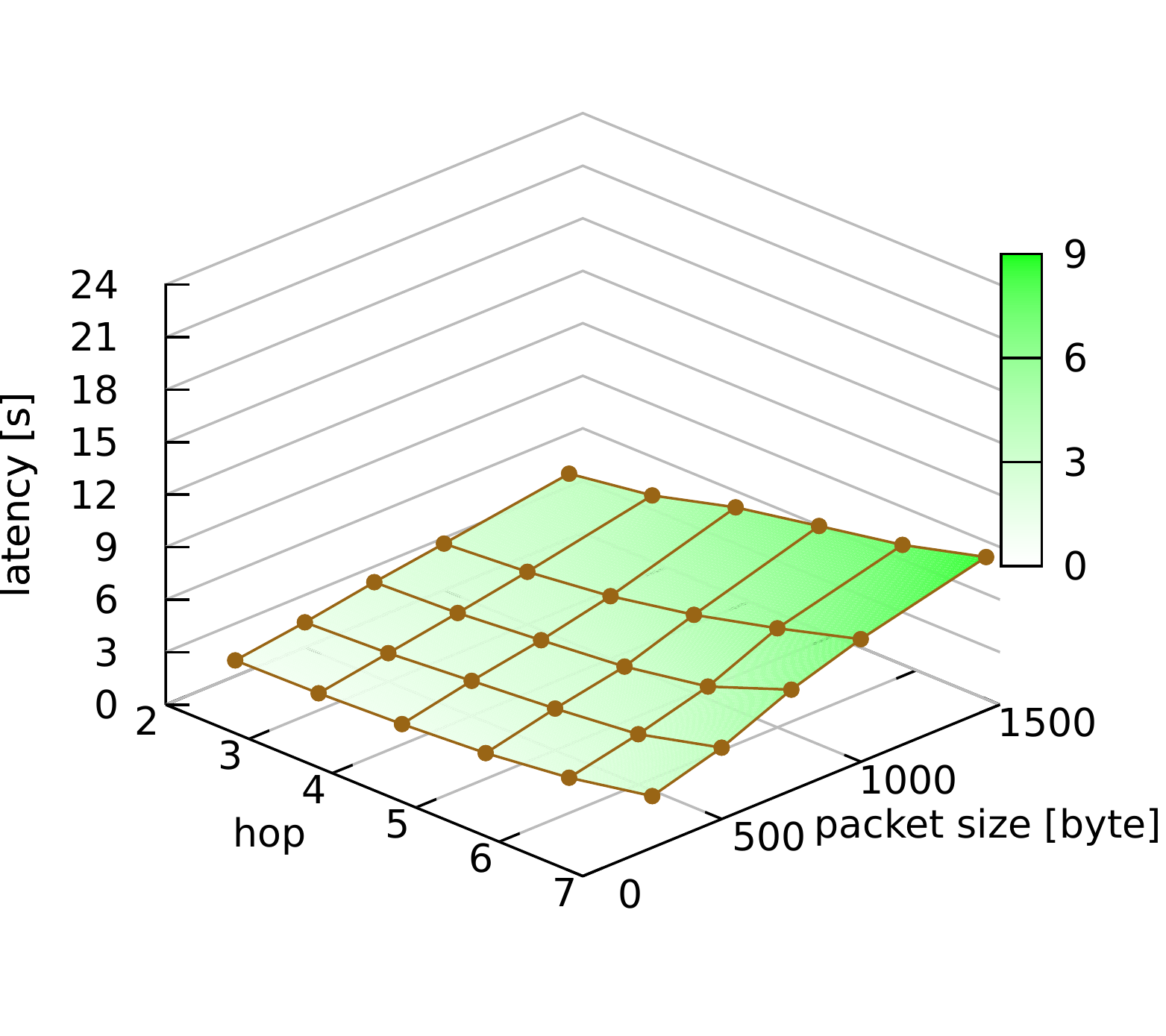}%
    \caption{$end-to-end$}
    \label{fig:3D_packet_size:e2e}
  \end{subfigure}
  \begin{subfigure}{.32\columnwidth}
    \includegraphics[trim=0cm 1.5cm 0cm 2cm,
    width=5.5cm]{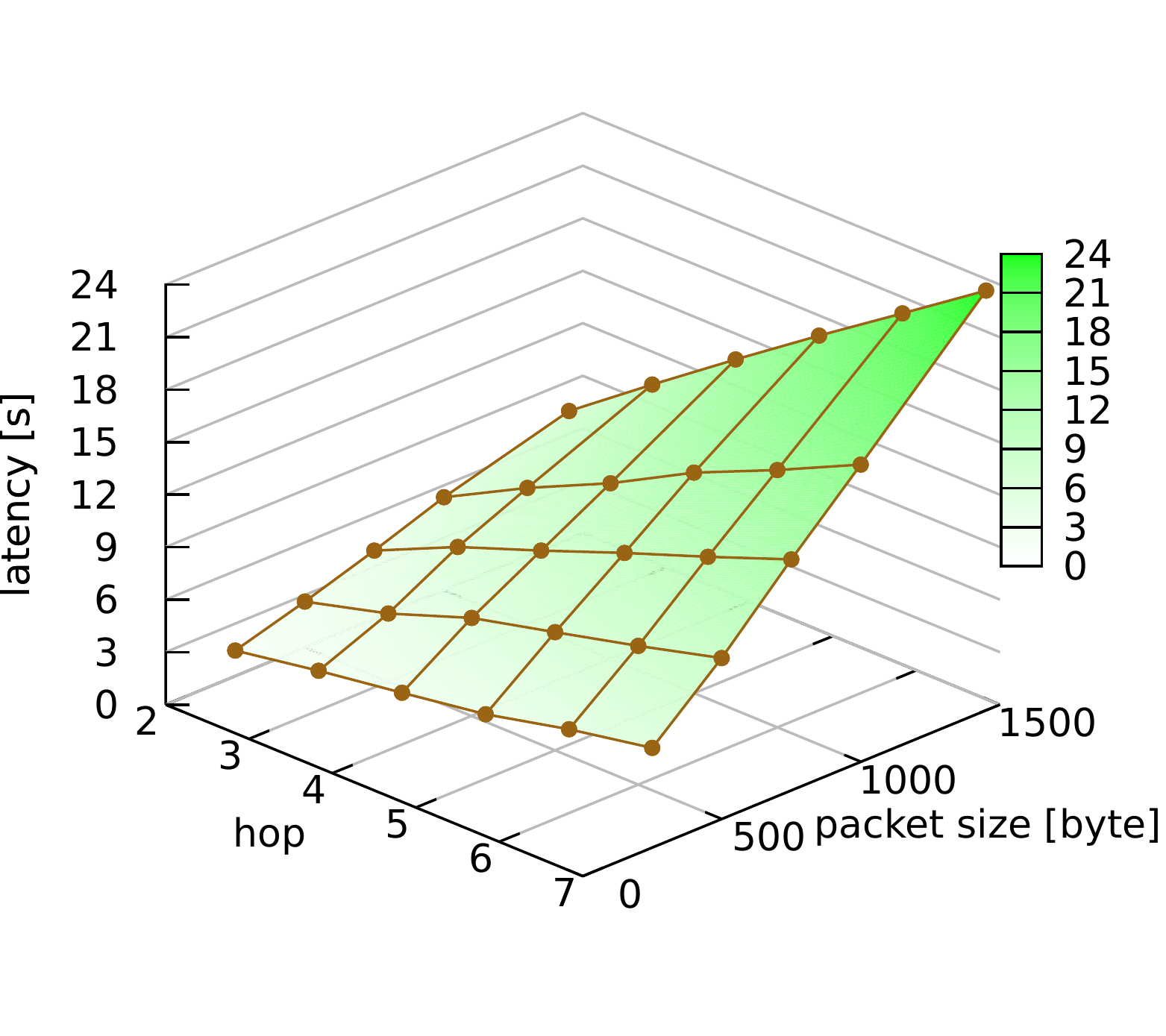}%
    \caption{$hop-by-hop$}%
    \label{fig:3D_packet_size:hbh}
  \end{subfigure}
  \begin{subfigure}{.32\columnwidth}
    \includegraphics[trim=0cm 1.5cm 0cm 2cm,
    width=5.5cm]{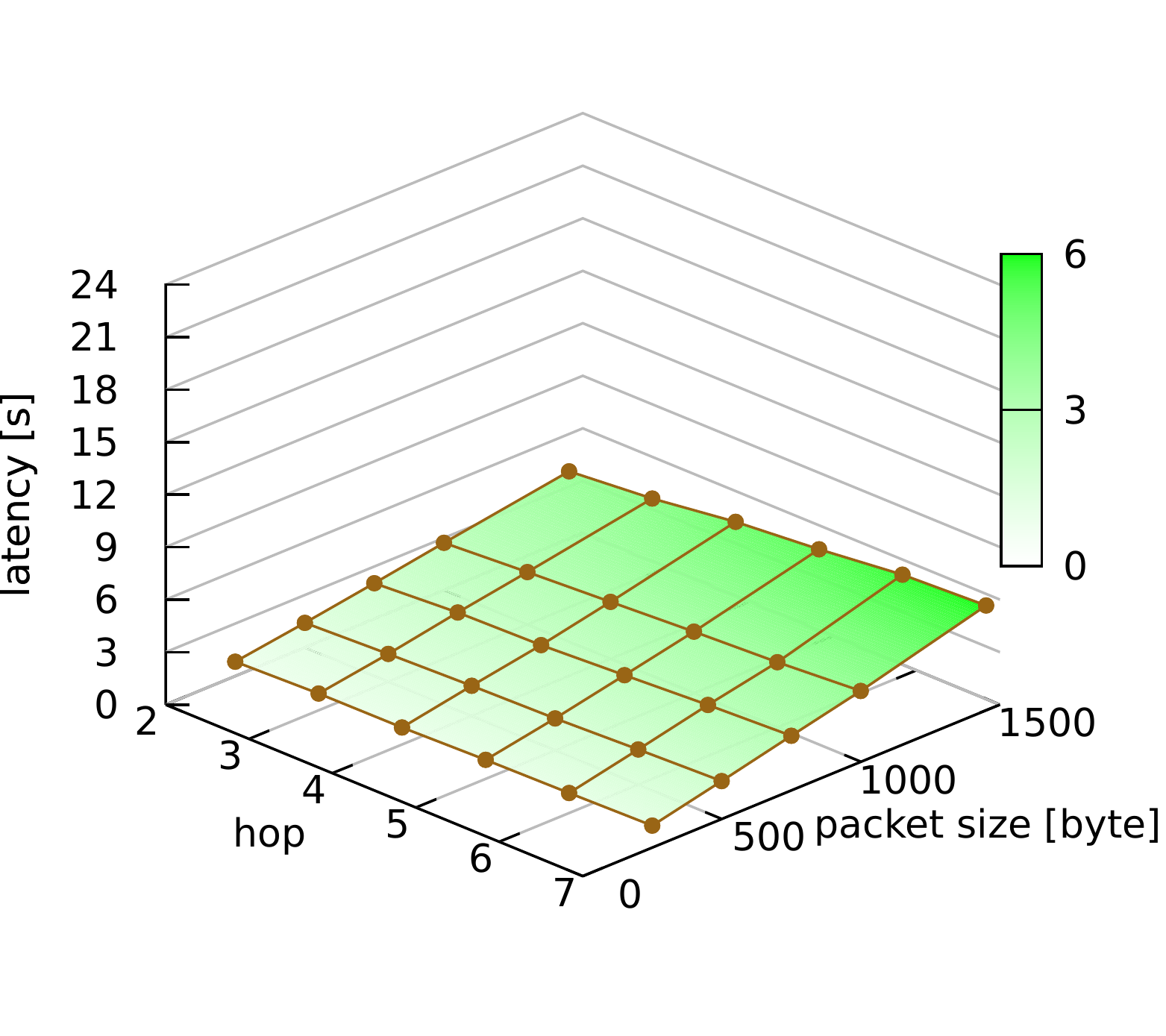}%
    \caption{$RLNC$}%
    \label{fig:3D_packet_size:rlnc}
  \end{subfigure}
  \label{figabc}
  \caption{Latency for the three transmission schemes depending on
    number of hops and packet size (Packets 64 - Loss 10\% - Bitrate
    0.25 Mb/s).}
  \label{fig:3D_packet_size}
\end{figure*}

In Fig.~\ref{fig:3D_losses} latency for the three transmission schemes depending on number of hops and loss probabilities is given. In the case of small number of hops with low loss E2E can keep pace with RLNC, at the expense of more sent packets. However, the latency increases significantly for large number of hops that are highly error prone. For HbH it increases linearly with the number of hops and increases with the probability of losses as given in Equation~\ref{delay:hbh}. RLNC has a lower latency than the other two schemes over a wide range of parameters.

In Fig.~\ref{fig:3D_gain} the gain of RLNC over the two schemes are given, namely E2E versus RLNC and HbH versus RLNC in Fig.~\ref{fig:3D_gain:e2e} and Fig.~\ref{fig:3D_gain:hbh}, respectively for a better comparison of the three schemes. The gain is calculated by the division of the latency either for E2E or HbH and RLNC. The plots in Fig.~\ref{fig:3D_gain} show a clear gain of RLNC over the other schemes. The maximum gain over E2E and HbH for the given parameters is 16x and 6x, respectively. Note, in Fig.~\ref{fig:3D_gain:hbh} the axes have been switched in order to increase visibility.

In Fig.~\ref{fig:3D_packet_size} the loss probability is still set to $10\%$ and the latency is plotted against the number of hops and the packet size. Most communication scenarios will use the maximum transfer unit (MTU) size of {\raise.17ex\hbox{$\scriptstyle\sim$}}1500 bytes, but smaller packet size will probably come up in future. For all three forwarding schemes latency increases linearly with packet size but HbH suffers the most as it is slower up to 4 times -- and E2E is up to 1.5 times -- compared to RLNC. Considering that the rate of loss is small what really makes the difference here is the different packet forwarding mechanisms described in Fig.~\ref{fig:schemes}. So with very small losses E2E can operate almost as efficient as RLNC, because the few number of extra packets, but HbH still slow due to the decoding-encoding at the middle nodes. So summing up the cases observed RLNC does not resonate as much as the other two schemes and breaks new grounds for future networking systems.

\section{Conclusion and Future Work}
\label{sec:con}

In this paper we have investigated some of the most important advantages of RLNC, the modern form of network coding, and we have shed some light on its application possibilities. We have provided a detailed comparison of RLNC and other coding strategies in terms of latency and traffic imposed to the network. In order to facilitate the use of RLNC we have also presented a prototype architecture demonstrating a feasible integration in SDN environment by using Virtualized Networking Functions. The VNFs implement RLNC functionality that enables us to leave all the management and traffic steering tasks on the SDN controller. This lead to flexible and automated deployment of service chains comprising network coding specific features, hereby introducing the code centric networking in SDN environment which is at same time compatible with the traditional packet switched networks. 

According to our results - both analytical and measurement - RLNC not only outperforms the others, as efficiently decreasing latency and number of sent packets, but also introducing flexibility by enabling packet forwarding without a centralized scheduling logic. In the future we will extend the measurements and the SDN prototype with multi-path functionality, since multipath  communication will not only increase the throughout and resilience, but also contribute to a further decreased latency. The need for RLNC over other coding techniques will become even more evident in the multipath context. Instead of the FullRLNC mode of Kodo, in the future we can use the sliding window approach that will reduce further the packet delay significantly.

\appendices
\section{Recursive formula for calculating latency in a two hop system}
\begin{figure*}[!t]
\centering
\begin{subfigure}{.32\columnwidth}
\includegraphics[trim=0cm 1.2cm 0cm 2.4cm, width=5.7cm]{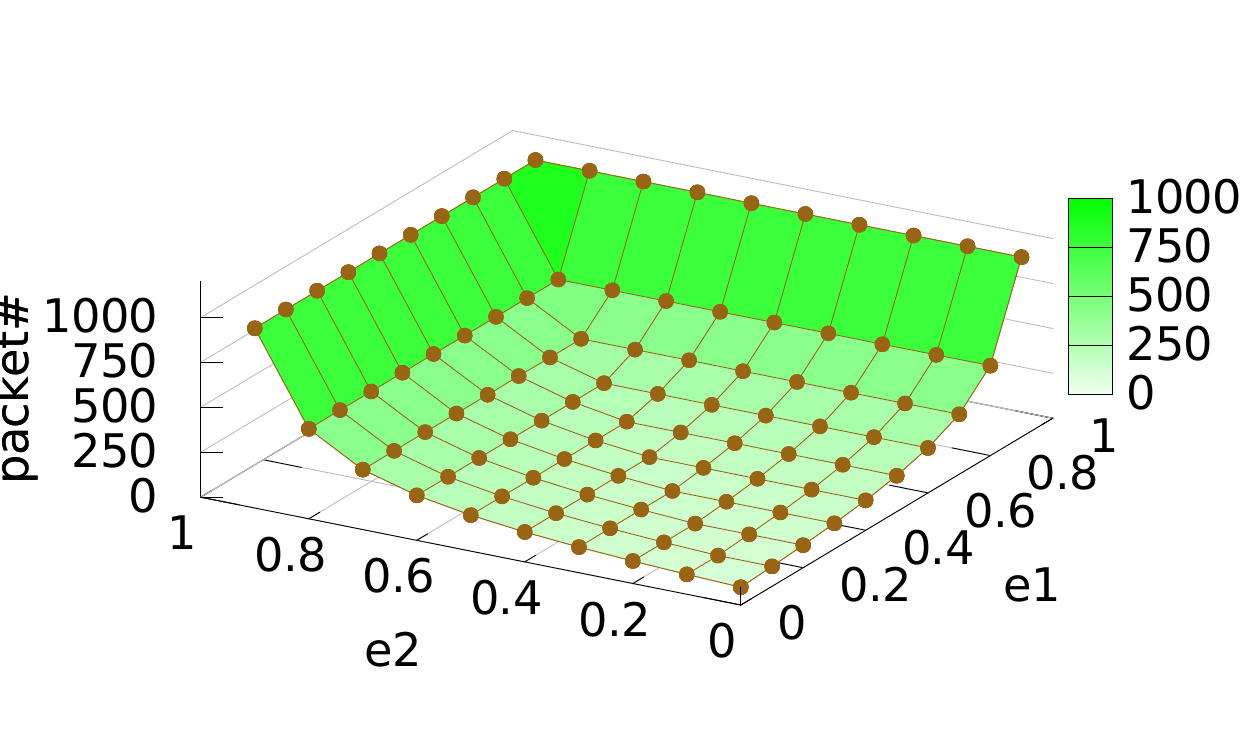}%
\caption{Latency calculated from theory in Eq. (5).}
\label{fig:3D_rlnc:theo2}
\end{subfigure}
\begin{subfigure}{.32\columnwidth}
\includegraphics[trim=0cm 1.2cm 0cm 2.4cm, width=5.7cm]{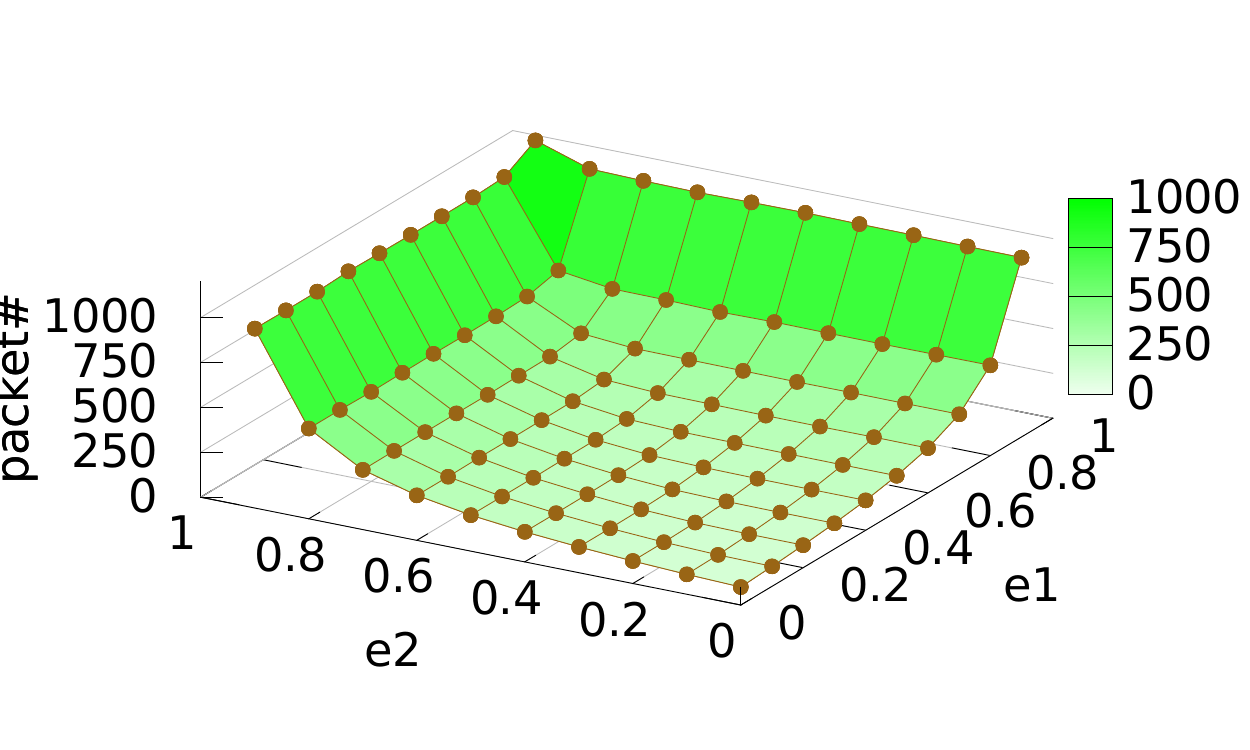}%
\caption{Latency calculated from simulations.}%
\label{fig:3D_rlnc:sim2}
\end{subfigure}
\begin{subfigure}{.32\columnwidth}
\includegraphics[trim=0cm 1.2cm 0cm 2.4cm,width=5.7cm]{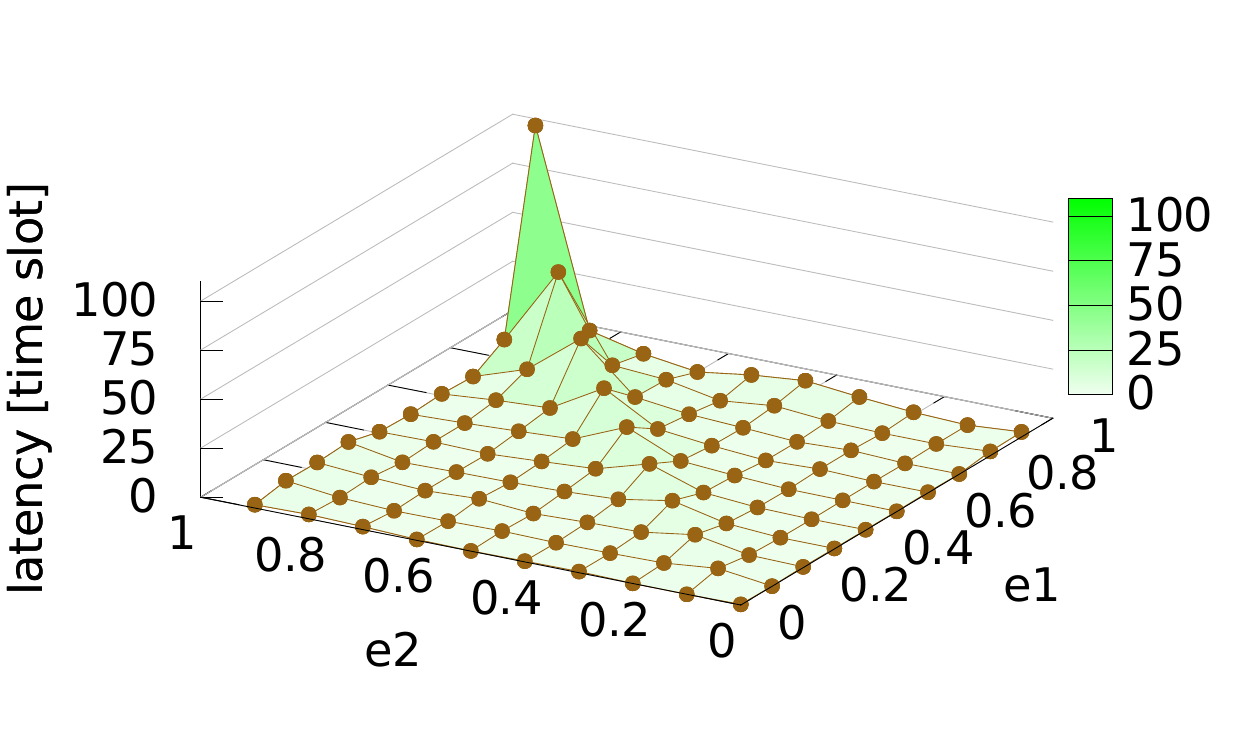}%
\caption{The difference between the theory in Eq. (5) and the simulations.}%
\label{fig:3D_rlnc:diff2}
\end{subfigure}
\caption{Number of time slots required to successfully send one generation of packets using RLNC through a two channel network with loss probabilities of $\epsilon_1$ and $\epsilon_2$.}
\label{fig:3D_rlnc2}
\end{figure*}

To take a closer look on the difference between measurements and theory in Fig.~\ref{fig:theory_vs_sim:latency:fifty} for RLNC we created a simulation environment for the most simple case of RLNC comprising only one encoder, one recoder and one decoder with two error prone links between them. For the sake of simplicity, in the simulation we calculated link delay as zero and used time slots as an analogy for inter packet time, so each node sends one packet per slot. Fig.~\ref{fig:3D_rlnc2} shows the latency as number of time slots required until the full message was decoded in dependency of the two channel error probabilities $\epsilon_1$ and $\epsilon_2$. It can be seen that difference occurs between theory and simulation only when error probabilities are close to each other ($\epsilon_1 \approx \epsilon_2$), which suggests that Eq.~\ref{delay:rlnc} isn't precise but what actually happens is that due to the finite generation sizes the actual loss on the two channels during the simulation is not exactly $\epsilon_1$ and $\epsilon_2$ but slightly differs from them as a random variable. When the error rate on one channel is significantly higher than the rate on the other the impact of this effect is very small since there is very low chance that more packet will be lost on the lower error rate channel (we can say that the higher error suppresses the lower). When the two error rates are close to each other we always have to calculate with the higher random value thus the result will be higher than expected from the theory. 

\begin{figure}[t]
\centering
\begin{subfigure}{.4\columnwidth}
\includegraphics[height=.65\linewidth]{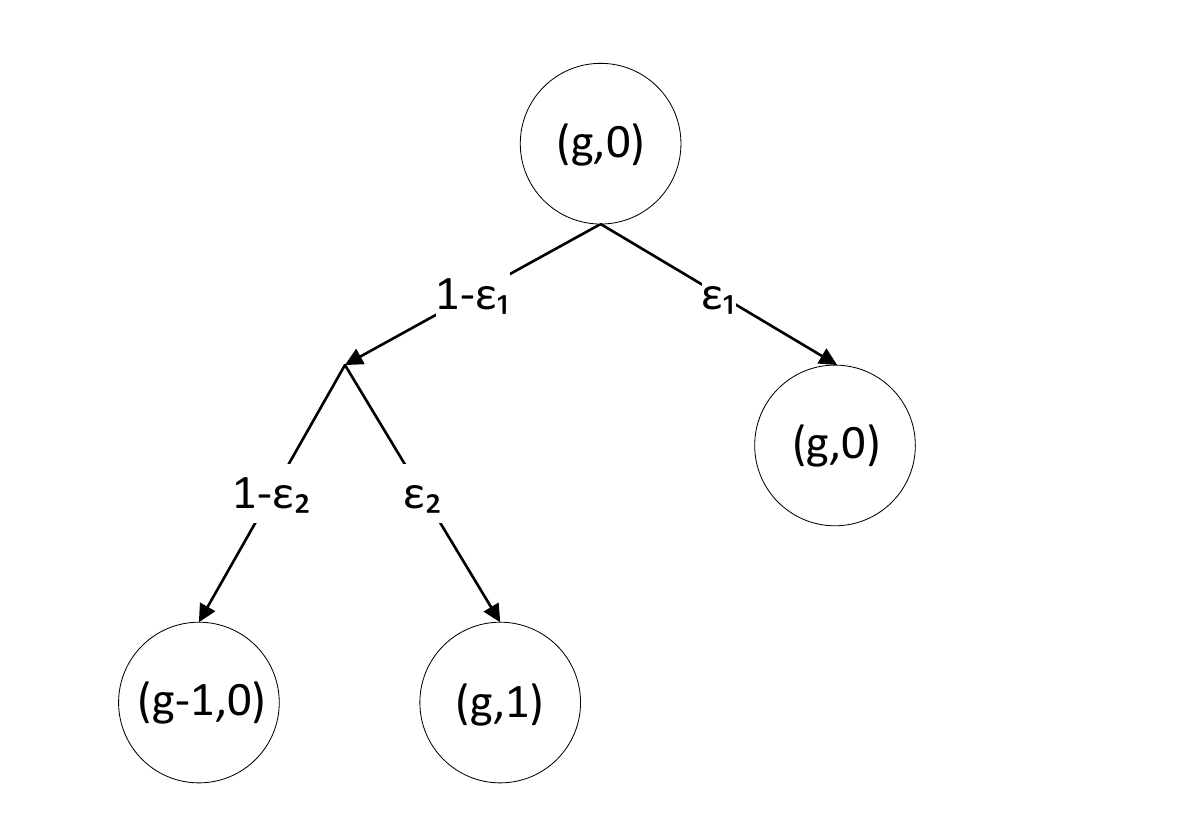}%
\caption{State transition graph when the recoder is empty.}
\label{fig:forwarding:a}
\end{subfigure}
\begin{subfigure}{.4\columnwidth}
\includegraphics[height=.65\linewidth]{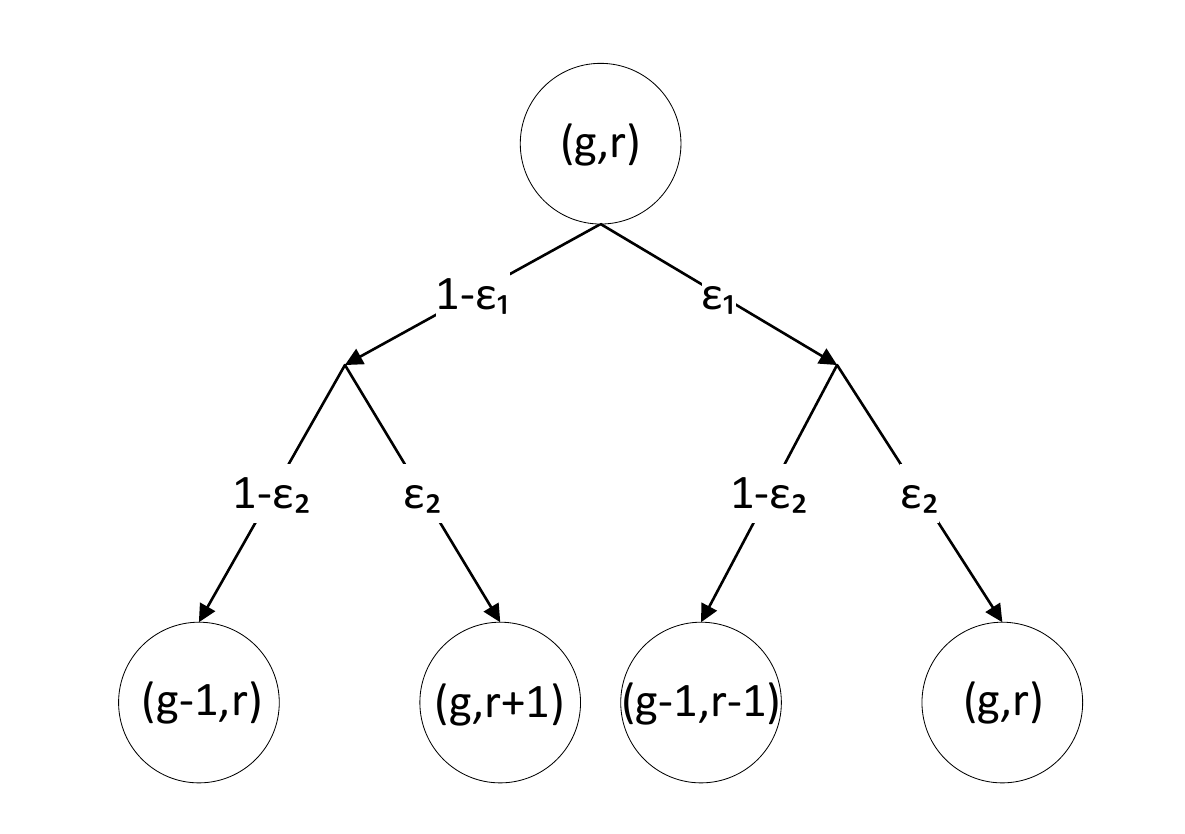}%
\caption{State transition graph when the recoder has at least on linearly independent packet.}
\label{fig:forwarding:b}
\end{subfigure}
\caption{Graphical representation of a two hop system: $g$ is the number of packets that the decoder still needs in order to decode the full generation, $r$ is the number of linearly independent packets in the recoder.}
\label{fig:forwarding}
\end{figure}

\begin{figure*}[!t]
\centering
\begin{subfigure}{.32\columnwidth}
\includegraphics[trim=0cm 1.2cm 0cm 2.4cm, width=5.5cm]{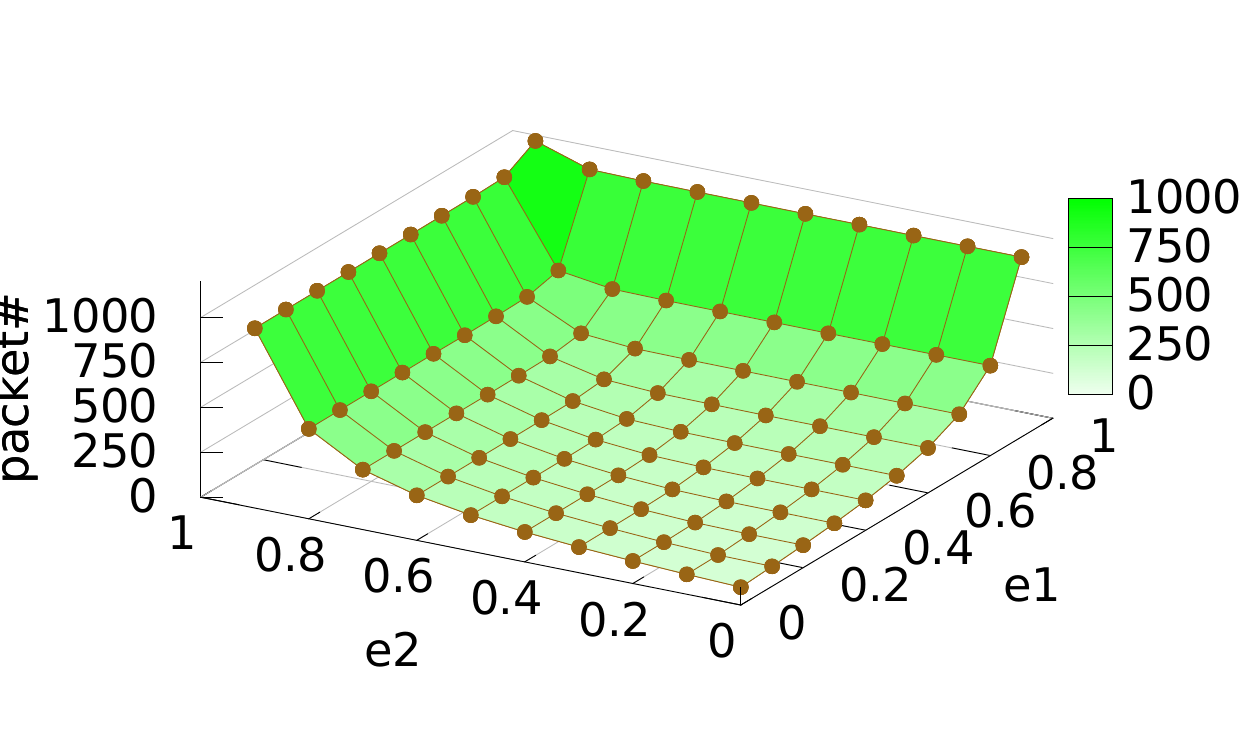}%
\caption{Latency calculated from recursive formula in Eq. (7).}%
\label{fig:3D_rlnc:theo}
\end{subfigure}
\begin{subfigure}{.32\columnwidth}
\includegraphics[trim=0cm 1.2cm 0cm 2.4cm, width=5.5cm]{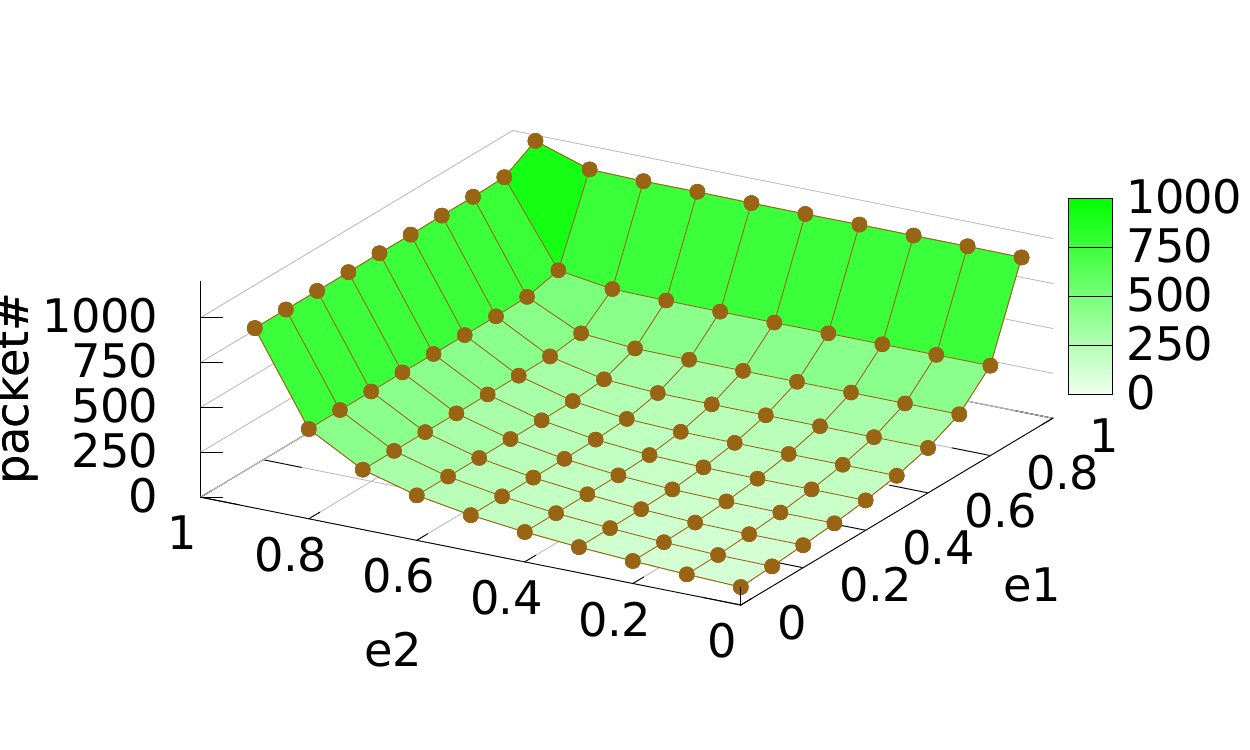}%
\caption{Latency calculated from simulations.}
\label{fig:3D_rlnc:sim}
\end{subfigure}
\begin{subfigure}{.32\columnwidth}
\includegraphics[trim=0cm 1.2cm 0cm 2.4cm,width=5.5cm]{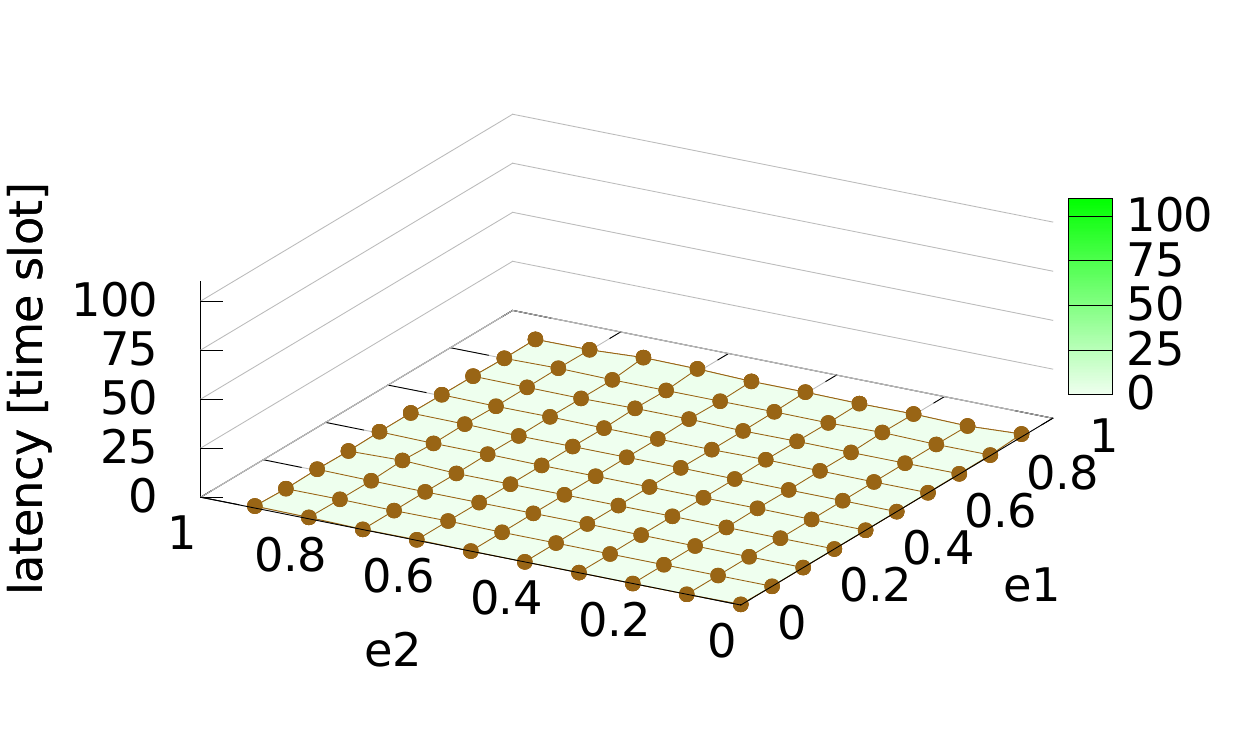}%
\caption{The difference between the values calculated from the recursive formula in Eq. (7) and the simulations.}%
\label{fig:3D_rlnc:diff}
\end{subfigure}
\caption{Number of time slots required to successfully send one generation of packets using RLNC through a two channel network with loss probabilities of $\epsilon_1$ and $\epsilon_2$.}
\label{fig:3D_rlnc}
\end{figure*}

To check our intuition we created a recursive formula based on the forwarding process that can calculate the latency in this 2 hop case. Since forwarding ends when the full generation comprising $G$ packets is delivered to the decoder, let $(g,r)$ be the state of the system, where $g$ is the number of packets that has to be delivered to the decoder and $r$ the number of linear independent -- i.e. the useful -- packets the recoder has and can send innovative packets based on them. This let us to distinguish two fundamentally different situations  (Fig.~\ref{fig:forwarding}), $(i)$ when recoder is running dry (Fig.~\ref{fig:forwarding:a}) and $(ii)$ when the recoder has useful packets to forward an innovative packet (Fig.~\ref{fig:forwarding:a}). When the recoder is running dry it means that if the packet from the encoder is lost the recoder can not send an innovative packet to the decoder thus the system stays in the same $(g,0)$ state but one time slot wasted. In~\cite{dikaliotis2009delay} authors present a similar model for delay calculation in RLNC and they state that a closed formula is too complex when $g$ is larger than 4. After that they use a random variable for modelling the delay thus -due to our knowledge- we are the first ones to present a closed recursive formula for delay in RLNC using arbitrary generation number and link losses in a two hop network.

\label{appendix}
We differentiate three cases as follows.
\begin{enumerate}
 \item If $r = g$: in case the recorer has enough number of linearly independent packets to send the reminder of the generation to the decoder, it does not need any extra packet from the encoder. Thus it means that we have to send $g$ number of packet through  a single link with an error probability of $\epsilon_2$. In this case the expected number of time slots can be calculated as follows. 
  \begin{equation}
   E(g,g) = g \cdot \frac{1}{1­-\epsilon_2}
  \end{equation}
 \item If $r = 0$: in this situation if the packet from the encoder get lost the recoder can not send an innovative packet to the controller (see Fig.~\ref{fig:forwarding:a}). The expected number of time slot needed for sending the remaining packet can be calculated with the following recursive formula.
\setlength{\arraycolsep}{0.0em} 
\begin{IEEEeqnarray}{rCl}
   E(g,0) & = & 1 + (1­-\epsilon_1) (1­-\epsilon_2) E(g­-1,0) \nonumber \\
   & & {+} (1­-\epsilon_1) \epsilon_2 E(g,1) + \epsilon_1 E(g,0) \nonumber \\
   & = & \frac{1}{1-\epsilon_1} + (1­-\epsilon_2) E(g-­1,0) + \epsilon_2 E(g,1)\nonumber \\
   & & 
\end{IEEEeqnarray}
 \item If $0 < r < g$: in this case the recoder already has $r$ number of linearly independent packets thus even when the packet from the encoder to the recoder get lost the recoder can send an innovative packet to the decoder (see Fig.~\ref{fig:forwarding:b}). The expected number of time slot needed for sending the remaining packet can be calculated with the following recursive formula.
\begin{IEEEeqnarray}{rCl}
   E(g,r) & = & 1 + (1­-\epsilon_1) (1-­\epsilon_2) E(g­-1,r)  \nonumber \\
   & & {+} (1-­\epsilon_1) \epsilon_2 E(g,r+1) \nonumber \\
   & & {+} \epsilon_1 (1-­\epsilon_2) E(g-­1,r-­1)  \nonumber \\
   & & {+} \epsilon_1 \epsilon_2 E(g,r) \nonumber \\
   & = & \frac{1}{1-­\epsilon_1 \epsilon_2} \nonumber \\
   & & {+} \frac{(1­-\epsilon_1) (1-­\epsilon_2)}{1-­\epsilon_1 \epsilon_2} E(g­-1,r)  \nonumber \\
   & & {+} \frac{(1-­\epsilon_1) \epsilon_2}{1­-\epsilon_1 \epsilon_2} E(g,r+1)  \nonumber \\
   & & {+} \frac{\epsilon_1 (1-­\epsilon_2)}{1­-\epsilon_1 \epsilon_2} E(g­-1,r-­1) 
\end{IEEEeqnarray}
\end{enumerate}
\setlength{\arraycolsep}{5pt}

On Fig.~\ref{fig:3D_rlnc} we compared again the simulations with the values derived from the recursive formula and we get almost no difference, so this formula describes the process precisely. However, generalizing this recursive formula for $H$ number of hops has exponential complexity since the state system can be described by $H$ number of variables (the number of packets that the decoder still needs to decode the full generation and the linearly independent packets in every $H-1$ recoders).

\bibliographystyle{IEEEtran}
\bibliography{ref}

\end{document}